\documentclass[%
    reprint,
    showpacs,preprintnumbers,
    showkeys,
    amsmath,amssymb,
    aps,
    prb,
    floatfix,
    longbibliography,
	nobalancelastpage,
]{revtex4-2}

\usepackage[english]{babel}
\usepackage[utf8]{inputenc}
\usepackage{graphicx}
\usepackage{bm}
\usepackage{hyperref}
\usepackage{xcolor}
\usepackage{epsfig}
\usepackage{amsmath}
\usepackage{amsfonts}
\usepackage{amssymb}
\usepackage{amsthm}
\usepackage{tikz}
\usepackage{dsfont}
\usepackage[all]{nowidow}
\usepackage{pifont}
\usepackage{algorithm}
\usepackage{algpseudocode}
\usepackage[capitalise]{cleveref}
\usepackage{nicefrac}
\usepackage{units}
\usepackage{mathtools}
\usepackage{algorithm}
\usepackage{algpseudocode}
\usepackage{tabularx}

\let\doi\undefined
\usepackage{doi}

\usepackage[letterspace=130]{microtype}

\definecolor{refColor}{HTML}{0376E9}
\definecolor{figColor}{HTML}{E90303}
\definecolor{urlColor}{HTML}{0376E9}

\hypersetup{%
    unicode=false,                 
    pdftoolbar=true,               
    pdfmenubar=true,               
    pdffitwindow=false,            
    pdfstartview={FitH},           
    pdftitle={},                   
    pdfauthor={Nicolai Lang},      
    pdfsubject={Quantum physics},  
    pdfcreator={Nicolai Lang},     
    pdfproducer={Nicolai Lang},    
    pdfkeywords={},                
    pdfnewwindow=true,             
    colorlinks=true,               
    linkcolor=figColor,            
    citecolor=refColor,            
    filecolor=magenta,             
    urlcolor=urlColor              
}

\newcommand{\Ket}[1]{\left|#1\right>}
\newcommand{\bra}[1]{\mathinner{\langle{#1}|}}
\newcommand{\ket}[1]{\mathinner{|{#1}\rangle}}

\renewcommand{\vec}[1]{\mathbf{#1}}

\renewcommand{\vec}[1]{\boldsymbol{#1}}

\newcommand{\sub}[1]{{\text{\tiny #1}}}

\newcommand{\flpat}{\tilde{E}^p}

\newcommand{\CaptionMark}[1]{\textbf{#1}}
\newcommand{\lbl}[1]{(#1)}

\begin{document}

\title{Decoding the Projective Transverse Field Ising Model}

\author{Felix Roser}
\email{felix.roser@itp3.uni-stuttgart.de}

\author{Hans Peter Büchler}

\author{Nicolai Lang}

\affiliation{%
    Institute for Theoretical Physics III and Center for Integrated Quantum Science and Technology,\\
    University of Stuttgart, 70550 Stuttgart, Germany
}

\date{\today}


\begin{abstract}

	The competition between non-commuting projective measurements in discrete quantum circuits can give rise to entanglement transitions. It separates a regime where initially stored quantum information survives the time evolution from a regime where the measurements destroy the quantum information. Here we study one such system---the projective transverse field Ising model---with a focus on its capabilities as a quantum error correction code. The idea is to interpret one type of measurement as an error and the other type as a syndrome measurement. We demonstrate that there is a finite threshold below which quantum information encoded in an initially entangled state can be retrieved reliably. In particular, we implement the maximum likelihood decoder to demonstrate that the error correction threshold is distinct from the entanglement transition. This implies that there is a finite regime where quantum information is protected by the projective dynamics, but cannot be retrieved by using syndrome measurements.
\end{abstract}

\maketitle

\section{Introduction}
\label{sec:introduction}

Entanglement lies at the heart of quantum mechanics and plays a key
role in various fields, two important examples being \emph{quantum
error correction}~\cite{Terhal2015} and \emph{entanglement
transitions}~\cite{Skinner_2019}. Because of decoherence, quantum error
correction has been identified as an indispensable step towards scalable,
universal quantum computation~\cite{Knill_1996,Aharonov_1997}, and the
existence of quantum codes demonstrated by Shor~\cite{Shor_1994} is the bedrock
on which the promises of quantum computation rest~\cite{Shor_1994}. As it
turns out, entanglement between the physical qubits of a quantum code is a
necessary ingredient for quantum error correction.
Independent of these considerations, the notion of
\textit{monitored quantum circuits} has gained traction in recent
years~\cite{Li_2018,Skinner_2019,Chan_2019,Szyniszewski_2019,Li_2019,Gullans_2020,Tang_2020,Jian_2020,Turkeshi_2020,Bao2020}.
The idea is to study quantum circuits built
from random unitary gates, which typically lead to volume-law entangled
states, and monitor them with local, projective measurements. The
latter counter the entanglement growth and their competition
can lead to transitions in the entanglement structure of the
system.
Strikingly, there are regimes in which a finite rate of projective measurements
can be tolerated while still preserving long-range entanglement~\cite{Li_2018,Chan_2019,Skinner_2019}. It then seems
natural to study the error correction capabilities of such random systems.
In this paper, we study a specific model that has been shown to feature an
entanglement transition, and view it as a quantum code to illuminate the
relation between quantum error correction and entanglement.

The rationale of quantum error correction is the following~\cite{Nielsen_2010}:
First, the quantum information to be protected (the \emph{logical qubits}) is
mapped to a subspace of the full system in such a way that local operations
on few physical qubits can neither leak information about the amplitudes
nor change them. This subspace is known as \emph{code space} and necessarily
comprises states that share entanglement between the physical qubits. If such
a system is projectively monitored by the environment, two things can happen:
Either the environment gains access to the amplitudes and they are lost
irretrievably or it does not and the amplitudes are still hidden in the state
of the system. However, as the interaction with the environment injects entropy
into the system, these hidden amplitudes cannot be accessed right away. The
next step of quantum error correction is therefore to measure observables
that do not destroy the encoded amplitudes but retrieve information about
the errors induced by the environment, thereby lowering the entropy of the
system. The extracted information is called \emph{syndrome} and can be used
to reconstruct the unknown errors to access the hidden amplitudes; this last
step is referred to as \emph{decoding} and the method used to convert the
syndrome into a tentative error pattern is the \emph{decoding algorithm}. One
can use different decoding algorithms for the same quantum code. Whether
decoding for a particular instance succeeds depends on the error pattern,
the knowledge of the decoder (the syndrome), and the decoder itself. The
maximum error rate up to which decoding with a given decoder succeeds on
average is known as \emph{error threshold}, a quantity that in many cases
can only be approximated numerically.
Prominent examples for quantum codes are Shor's
nine-qubit code~\cite{Shor_1995}, the seven-qubit Steane
code~\cite{Steane_1996}, and scalable codes derived
from topologically ordered systems like the toric/surface
codes~\cite{Bravyi_1998,Kitaev_2003,Freedman_2001,Dennis_2002}. For
all of these, efficient decoding algorithms
and numerical results for their error thresholds are
known~\cite{Paz-Silva_2010,Crow_2016,Fowler_2009,Wootton_2012,Bravyi_2014,Fowler_2012_ThresholdProof,Wang_2009};
the particularly high thresholds of surface codes make them promising
candidates for real-world implementations~\cite{Krinner_2022,Zhao_2021}.

The notion of monitored quantum circuits that give rise to entanglement
transitions was introduced in Refs.~\cite{Li_2018,Skinner_2019,Li_2019} and
originated from the question whether extensive, coherent subsystems can be
stabilized in the presence of environmental noise. Typically, one starts with
a product state of an extensive number of qubits equipped with a geometry
and applies a random sequence of 2-local unitaries (to model coherent,
local interactions) interspersed with single-qubit projective measurements;
the relative rate of unitaries and projective measurements is the parameter
of the system. The unitaries are ideally drawn from the Haar measure,
but often restricted to the Clifford group instead to make use of the
stabilizer formalism for efficient simulation~\cite{Nielsen_2010,
Gottesman_1996,Gottesman_1997,Gottesman_1998_Theory,
Gottesman_1998,Aaronson_2004}.
The findings of Refs.~\cite{Li_2018,Skinner_2019,Li_2019,Chan_2019} show
that there is a finite, critical rate for the projective measurements at
which a continuous transition in the entanglement structure takes place:
below the critical rate, the unitary evolution dominates and stabilizes the
system in a volume-law phase, whereas, above the critical rate, the projective
measurements prevail so no long-range entanglement can build up (indicated
by area-law states with only local entanglement). Later, it was shown that
similar transitions can be found in systems where the unitaries are replaced by
projective measurements that do not commute with the projective measurements of
the environment~\cite{Nahum2020,Lang_2020,Ippoliti2021,Lavasani2021,Sang2021}.
Depending on the measurements~\cite{Ippoliti2021}, the entanglement transitions
in such purely projective models often separate phases of different area laws
(instead of volume-law and area-law phases).
One such model, derived from the transverse field Ising model and dubbed
\emph{projective transverse field Ising model (PTIM)} was studied in
Ref.~\cite{Lang_2020}. It features a critical monitoring rate dictated by
bond percolation. Below this rate, the system exhibits long-range entangled
states with an area law; above, it transitions into product states with
short-range entanglement. Here we use this model to study the connection
between quantum error correction and entanglement transitions.

That these two fields are related is not hard to see: If one initializes
a monitored quantum circuit not in a product state but an entangled
state that encodes the amplitudes of a logical qubit nonlocally, one
can ask how long these amplitudes survive the evolution of the system
and how the average lifetime scales with the system size. It turns
out that this is an alternative characterization of the entanglement
transition~\cite{Aharonov_2000,Choi_2020,Gullans_2020,Lang_2020,Li_2021_2,Fan2021,Nakata2021,Gullans2021,Fidkowski2021}: below
the critical monitoring rate (in the entangling phase), the lifetime grows
exponentially with the system size so the amplitudes survive indefinitely
in macroscopic systems; by contrast, the growth is subexponential (typically
logarithmic) in the disentangling phase.
In this sense, monitored quantum circuits in the entangling phase can be
seen as random quantum error correction codes that protect amplitudes
by scrambling them into distributed degrees of freedom faster than the
environment can extract information. This relates to the first (encoding)
stage of quantum error correction explained above. What is not so clear is
the second (decoding) stage: Under which conditions and how is it possible
to \emph{retrieve} the scrambled amplitudes? As for quantum codes, one may
expect an error threshold for the projective measurements of the environment
below which retrieval is possible by some decoding algorithm. However, it is
unclear how the error threshold relates to the entanglement threshold except
that the latter poses an upper bound on the former (in the disentangling
phase, the amplitudes are lost and cannot be retrieved).

In this paper, we answer these questions specifically for the one-dimensional
PTIM with a detailed study of several
decoding algorithms.
We analyze the decoders numerically making use of a mapping between
trajectories with projective measurements and ``classical'' trajectories
with unitary operations. We introduce quantities to evaluate the performance
of decoders quantitatively. Our analysis includes a na\"ive decoder based on
\emph{majority voting} [the maximum likelihood decoder (MLD) for classical repetition
codes], which is an instructive approach that, unfortunately, fails. We
then discuss and evaluate a decoder based on \emph{minimum weight perfect
matching} (MWPM), an algorithm that is often used for the decoding of topological
surface codes. The algorithm has been applied successfully to the decoding of
the PTIM by Li and Fisher~\cite{Li_2021};
we verify their results qualitatively and find a finite error threshold
below which the decoder reliably retrieves the encoded quantum information.
We then extend these results and implement the MLD of this system, i.e., the provably optimal decoder. We show that its
decoding threshold provides only a marginal improvement over the MWPM decoder. Furthermore, we show that the decoding threshold of
the MLD is distinct from the entanglement transition
of the PTIM. We conclude that there is
a finite range of parameters where the encoded amplitudes \emph{survive}
the monitoring by the environment but cannot be \emph{retrieved} without
having access to the full system dynamics.

The remainder of this paper is structured as follows. In \cref{sec:setting},
we start with a description of the model and focus on its interpretation
as a quantum code. We continue by formalizing the task of decoding and
introduce quantities to gauge the performance of decoders. In \cref{sec:MVD},
we present a brief discussion of a na\"ive decoder based on majority voting
and demonstrate why it fails. Taking guidance from this failure, we construct
a decoder based on MWPM in \cref{sec:MWPM} and
study its performance numerically; we find a finite error threshold. In
\cref{sec:MLD}, we implement the MLD, which we
find to perform slightly better than the MWPM
decoder. We conclude in \cref{sec:summary} with a summary and outlook. In
\cref{app:simulateClassical,app:MLDProof,sec:pseudoclassicalSampling,app:finitesize}
we provide technical details and proofs for some claims in the main text.


\section{Setting}
\label{sec:setting}

\subsection{The model}

\begin{figure*}[tbp]
	\centering
	\includegraphics[width=0.80\textwidth]{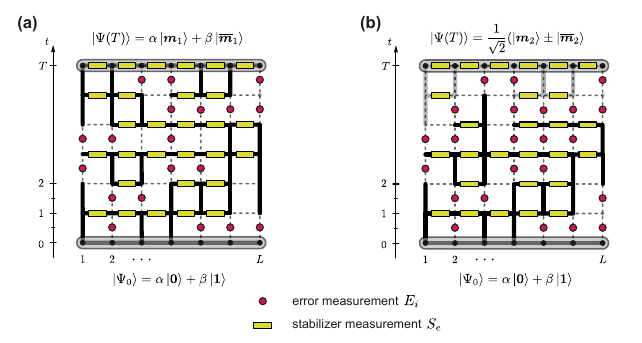}
	\caption{%
        \CaptionMark{The model.} 
        The projective transverse Ising model (PTIM) is a stochastic process
        where in each time step $t\mapsto t+1$ first error measurements
        $E_i=\sigma_i^x$ occur with probability $p$ per site, and then syndrome
        measurements occur with probability $1-q$ per edge $e=(i,i+1)$ along
        a one-dimensional open chain of $L$ qubits. In this paper, we study
        sample averages over many trajectories $\mathcal{T}=\{\Psi(t)\}$
        for initial states $\ket{\Psi(t=0)}=\ket{\Psi_0}=\alpha\ket{0\dots
        0}+\beta\ket{1\dots1}$ that belong to the code space of the quantum
        repetition code. We are interested in whether the logical amplitudes
        $\alpha,\beta\in\mathbb{C}$ survive the PTIM evolution for $t=T\sim L$
        time steps. To test this, we finalize each trajectory with a complete
        syndrome measurement. The final state $\ket{\Psi(T)}$ can then have
        only two forms:
        \lbl{a}~Trajectory where the amplitudes survive, which corresponds
        to the bond percolation of the initial cluster highlighted in
        black. While horizontal edges are active if there is a stabilizer
        measurement, vertical edges are active if there is \emph{no} error
        measurement. The cluster connected to the initial state at $t=0$
        carries the encoded amplitudes.
        \lbl{b}~Trajectory where the initial cluster does not percolate. In
        this case, the encoded amplitudes are lost. The final qubit patterns
        $\vec m_i$ depend on the trajectory $\mathcal{T}$.
    }
	\label{fig:setting}
\end{figure*}

We start by introducing the PTIM
and its interpretation as a quantum code. Consider an open chain of $L$
qubits prepared in the initial state $\ket{\Psi(t=0)}=\ket{\Psi_0}$. To
propagate the system from time step $t$ to $t+1$, we first loop through all
sites and measure with probability $p\in [0,1]$ the operator
\begin{equation}
	E_i=\sigma_i^x\,.
\end{equation}
Here, $\sigma_i^\alpha$ denotes the Pauli matrix $\alpha=x,y,z$ acting
on qubit $i=1,\dots,L$. We will refer to these measurements as \emph{error
measurements}; they describe the monitoring of the system by the environment.
Next, we loop through all edges $e=(i,i+1)$ between adjacent sites and measure
\begin{equation}
    S_e=\sigma_i^z\sigma_{i+1}^z
\end{equation}
with probability $1-q\in [0,1]$. We will refer to these as \emph{syndrome
measurements} as they are the stabilizers of the quantum repetition code (see
below). Note that $q\in [0,1]$ refers to the probability that a stabilizer
is \emph{not} measured.
Repeating this two-step process generates a trajectory
$\mathcal{T}=\{\Psi(t)\}$ of the PTIM, see \cref{fig:setting}~(a) for an
example. For a fixed initial state, the trajectory is uniquely determined by
the space-time patterns $E^p$ and $S^p$ of error- and syndrome measurements
and their respective measurement outcomes $E^r$ and $S^r$. With a slight
abuse of notation, we refer to a trajectory as $\mathcal{T}=(E,S)$ where
$E=(E^p,E^r)$ and $S=(S^p,S^r)$.
The parameters of this model are the probabilities $p$ for an error and $q$
for a missing stabilizer measurement,  and we are interested in properties of
this stochastic process when sampled over many trajectories for large system
sizes $L\rightarrow\infty$. Note that there are two types of randomness in
the model: the \emph{classical} randomness of measurement choices encoded
in the patterns $E^p$ and $S^p$, and the \emph{quantum} randomness from the
measurement outcomes $E^r$ and $S^r$.

For an initial product state of the form
$\ket{\Psi_0}=\ket{+}\otimes\dots\otimes\ket{+}\equiv\ket{+\dots +}$ (where
$\ket{+}$ denotes the state with $\sigma^x\ket{+}=\ket{+}$), it was shown in
Ref.~\cite{Lang_2020} that this process features an entanglement transition
at $p+q=1$ that is dictated by (anisotropic) bond percolation on the square
lattice: For $p+q>1$ (error and/or failed stabilizer measurements dominate)
the state $\Psi(t\rightarrow\infty)$ remains short-range entangled, whereas
for $p+q < 1$ (low error rate and/or syndrome measurements dominate), stable
long-range entanglement emerges.

By contrast, here we are interested in the error correction capabilities of
this model. To this end, we consider (typically entangled) initial states
of the form
\begin{align}
    \ket{\Psi_0}=\alpha\ket{0\dots 0}+\beta\ket{1\dots1},
    \label{eq:psi0}
\end{align}
with $\sigma^z\ket{0}=\ket{0}$ and $\sigma^z\ket{1}=-\ket{1}$;
$\alpha,\beta\in\mathbb{C}$ with $|\alpha|^2+|\beta|^2=1$ are the amplitudes
of the logical qubit to be protected. The two-dimensional subspace
$\mathcal{C}\ni\ket{\Psi_0}$ spanned by $\ket{\vec 0}\equiv\ket{0\dots 0}$
and $\ket{\vec 1}\equiv\ket{1\dots1}$ is the code space of the quantum
repetition code with stabilizer $\mathcal{S}=\langle\{S_e\}\rangle$, i.e.,
$\hat S\ket{\Psi}=\ket{\Psi}$ for all $\ket{\Psi}\in\mathcal{C}$ and $\hat
S\in\mathcal{S}$. This code can only correct bit flip errors $E_i=\sigma_i^x$
but not phase errors $\sigma_i^z$ (the latter correspond to parity-violating
terms in its fermionic representation; this is not of relevance in the
following).

To comply with the concept of quantum error correction, we modify the evolution
of the PTIM, initialized in \cref{eq:psi0}, by demanding that in the final
time step $t=T$ (where $T$ is fixed beforehand and typically $T\sim L$),
\emph{all} stabilizers $\{S_e\}$ are measured, irrespective of the failure
probability $q$. This forces the final state $\ket{\Psi(t=T)}$ into the
two-dimensional subspace $\mathcal{C}_{\vec{m}}$ spanned by $\ket{\vec m}$
and $\ket{\overline{\vec m}}=\prod_i\sigma_i^x\ket{\vec m}$ where $\vec
m = (m_1,\dots,m_L)$ with $m_i\in\{0,1\}$ is a qubit configuration that
corresponds to the syndrome measurements at $t=T$ and $\overline{\vec m}$
is its globally flipped configuration (which necessarily also matches the
syndrome). It was shown in Ref.~\cite{Lang_2020} that the final state has
only two possible forms,
\begin{align}
    \ket{\Psi(T)}=
    \begin{cases}
        &\tfrac{1}{\sqrt{2}}\ket{\vec m}\pm\tfrac{1}{\sqrt{2}}\ket{\overline{\vec m}}\\
        &\alpha\ket{\vec m}+\beta\ket{\overline{\vec m}},
    \end{cases}
\end{align}
where $\vec m=\vec m(\mathcal{T})$ is some qubit configuration
that depends on the trajectory $\mathcal{T}$. In the
first case, $\ket{\Psi(T)}=\tfrac{1}{\sqrt{2}}\ket{\vec
m}\pm\tfrac{1}{\sqrt{2}}\ket{\overline{\vec m}}$, the environment gained
access to the logical qubit through its measurements $E=(E^p,E^r)$ and no
recovery of the encoded amplitudes is possible. The probability for this
outcome approaches unity for $L,T\to\infty$ if $p+q > 1$, i.e., in the
disentangling phase of the PTIM. This is so because the cluster that carries
the amplitudes does not percolate, as shown in \cref{fig:setting}~(b).
In the second case, $\ket{\Psi(T)}=\alpha\ket{\vec m}+\beta\ket{\overline{\vec
m}}$, the cluster \emph{does} percolate [\cref{fig:setting}~(a)] and the
amplitudes survive the monitoring by the environment; however, they are now
encoded in a rotated basis $\ket{\vec m}=\hat C^\dag\ket{\vec 0}$, where $\hat
C = \prod_i (\sigma_i^x)^{C_i}$, $C=(C_1,\dots,C_L)\in\{0,1\}^L$, is a qubit
flip pattern that describes the effect of the measurements by stabilizers
and environment. The probability for this to happen approaches unity for
$L,T\to\infty$ in the entangling phase of the PTIM, i.e., for $p+q<1$.

The goal of this paper is to recover the encoded amplitudes from
$\ket{\Psi(T)}$, which is tantamount to finding a correction string
$C=C(\mathcal{T})$ such that
\begin{align}
    \hat C\ket{\Psi(T)}=\alpha\ket{\vec 0}+\beta\ket{\vec 1}=\ket{\Psi_0}\,.
\end{align}
We refer to this as \emph{decoding of the noisy quantum repetition code},
which we discuss in the next subsection.

\subsection{Decoding}
\label{subsec:decoding}

\begin{figure}[tbp]
	\centering
	\includegraphics[width=0.9\linewidth]{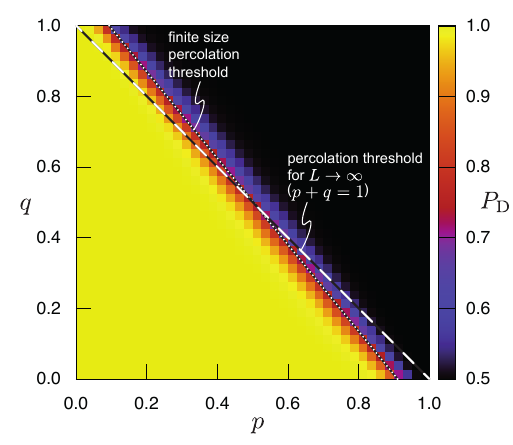}
	\caption{%
        \CaptionMark{Decoding probability with full knowledge.} 
        Probability $P_\sub{D}$ to successfully decode the PTIM with full
        knowledge of the error and stabilizer measurements $\mathcal{T}=(E,S)$
        as a function of the error probability $p$ and the stabilizer failure
        probability $q$ for $L=T=51$. For $p+q>1$, decoding fails almost
        surely because in the disentangling phase the amplitudes are lost due
        to the monitoring of the environment. Because of finite-size effects,
        the true percolation threshold is slightly tilted and smeared out. For
        each datapoint, we sampled $20\,000$ trajectories.
    }
	\label{fig:cluster}
\end{figure}

The entanglement transition at $p+q=1$ acts as an upper bound in the $(p,q)$
parameter space up to which decoding can possibly succeed, and we will focus
on $p+q < 1$ henceforth.
If we have access to the full trajectory $\mathcal{T}=(S,E)$, it is
straightforward to construct the decoding string $C$ deterministically because
we can simulate the evolution $\ket{\Psi(t)}$ efficiently using the stabilizer
formalism (there are even more efficient ways to achieve this, but this shall
not be our focus here). This is illustrated in \cref{fig:cluster} where we plot
the decoding probability $P_D$ as a function of the error probability $p$ and
the stabilizer failure probability $q$ (we will introduce the quantity $P_D$
formally below). Note that the threshold of the entanglement transition
is slightly tilted and smeared out due to finite-size effects; we checked
that for $L,T\to\infty$ the transition gets sharper and approaches the off-diagonal $p+q=1$ predicted by percolation theory.

\begin{figure*}[tbp]
	\centering
	\includegraphics[width=\textwidth]{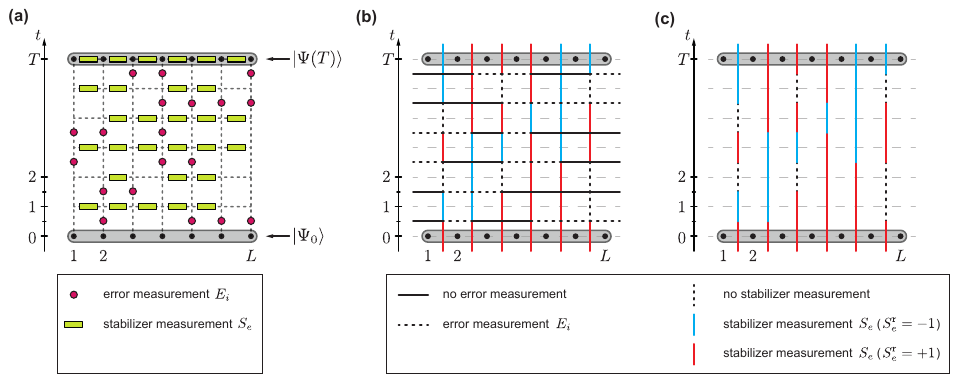}
	\caption{%
        \CaptionMark{Trajectory representations of the PTIM.} 
        \lbl{a}~Full trajectory of the PTIM reproduced from
        \cref{fig:setting}~(a) where the amplitudes survive the monitoring
        by the environment.
        \lbl{b}~Switching to the dual of the dashed space-time lattice
        in \lbl{a}, we can replace stabilizer measurements $S^p$ by solid
        \emph{vertical} edges, the colors of which encode the measurement
        outcomes $S^r$. \emph{Absent} error measurements are replaced by
        solid \emph{horizontal} edges. The emergent cluster structure is
        dual to the percolation cluster in \cref{fig:setting}~(a).
        This representation on the dual lattice will be useful later for constructing decoders.
        \lbl{c}~Decoders only have access to the syndrome data
        $S=(S^p,S^r)$, i.e., the vertical edges of the representation in
        \lbl{b}. \emph{Decoding} amounts to reconstructing the full pattern
        in \lbl{b} from the reduced one in \lbl{c} (or an equivalent one in
        a suitably defined sense).
    }
	\label{fig:system}
\end{figure*}

However, given our description of the measurements above, it is reasonable
to assume that we neither have access to \emph{when and where} error
measurements occurred ($E^p$) nor to their \emph{results} ($E^r$) as these
are performed by the environment; hence, we have only access to the syndrome
data $S=(S^p,S^r)$. In \cref{fig:system}, we illustrate the trajectory from
\cref{fig:setting}~(a) and construct an abstract, reduced representation
that illustrates the knowledge of the decoder.

This lack of knowledge makes the construction of $C$ a non-trivial problem.
However, the following observation helps us in this regard: Remember that the
syndrome data $S$ includes the outcomes of the full syndrome measurement
in the last step. As a consequence, we do know the rotated code space
$\mathcal{C}_{\vec{m}}$ spanned by $\ket{\vec m}$ and $\ket{\overline{\vec
m}}$. Given the amplitudes survived, the final state can have only two forms:
\begin{align}
    \ket{\Psi(T)}=
    \begin{cases}
        &\alpha\ket{\vec m}+\beta\ket{\overline{\vec m}}\,,\\
        &\alpha\ket{\overline{\vec m}}+\beta\ket{\vec m}\,.
    \end{cases}
    \label{eq:PsiTSurvived}
\end{align}
In the first case, the right choice for the correction string that defines
$\hat C$ is $C\equiv\vec m$, whereas, in the second case, it is $\overline
C\equiv\overline{\vec m}$. Decoding therefore amounts to choosing a decoding
string from a set $\{C,\overline C\}$ of two possible strings; selecting
the wrong one leads to a bit flip error on the logical qubit.

So, although we do not have access to an extensive amount of information,
namely, $E=(E^p,E^r)$, we are actually missing only a single bit, namely, whether
to choose $C$ or $\overline{C}$ to decode $\ket{\Psi(T)}$. A \emph{decoder}
or \emph{decoding algorithm} picks one of these two as a function of the
syndrome data $S$:
\begin{align}
    D\,:\,S=(S^p,S^r)\;\mapsto\,D(S)\in\{C,\overline C\}\,.
    \label{eq:D}
\end{align}
Note that in the following it is sufficient to start with an initial
state $\ket{\Psi_0}=\ket{\vec 0}$ to quantify the performance of the
decoder, as the system dynamics commutes with the logical $X$-operator
$X=\prod_i\sigma_i^x$. Thus, any initial state which is polarized in th
$X$ direction will never change its polarization, and it is best to choose
an initial state with vanishing $X$ polarization.

For a given trajectory $\mathcal{T}=(E,S)$  with final state
$\ket{\Psi_\text{f}(\mathcal{T})}\equiv\ket{\Psi(T)}$, we can define the
overlap
\begin{equation}
    f^\sub{qm}(\mathcal{T},C)
    :=\left|\bra{\Psi_0}\hat C\ket{\Psi_\text{f}(\mathcal{T})}\right|^2
    \label{eq:fqm}
\end{equation}
to identify the correct string $C$ to decode the quantum information. This
expression returns $1$ ($0$) for the correct (incorrect) string when
the amplitudes survive; in cases where the amplitudes are lost, one finds
$f^\sub{qm}=\tfrac{1}{2}$, irrespective of the chosen correction string.
This allows us to determine the performance of a decoder $D$. For a single
trajectory $\mathcal{T}=(E,S)$, we define
\begin{align}
    f^\sub{qm}_D(\mathcal{T})
    :=f^\sub{qm}(\mathcal{T},C=D(S))\,;
\end{align}
the performance of the decoder is then quantified by the probability to
correctly decode the quantum information,
\begin{align}
    P_D(p,q;L,T)
    :=\langle\langle f^\sub{qm}_D\rangle\rangle
    =\sum_{\mathcal{T}}P^\sub{qm}(\mathcal{T})\,f^\sub{qm}_D(\mathcal{T})\,,
\end{align}
where $\langle\langle\bullet\rangle\rangle$ denotes the sample average
over trajectories and $P^\sub{qm}(\mathcal{T})$ is the probability for the
trajectory $\mathcal{T}=(E,S)$. Note that $P^\sub{qm}(\mathcal{T})$ is a
highly nontrivial quantity: While the classical probability distribution that
governs $E^p$ and $S^p$ is straightforward to describe, the distributions of
measurement \emph{outcomes} $E^r$ and $S^r$ are not obvious as they depend
on the history of the trajectory and are not independent.

The quantity $P_D$ is our main figure of merit to evaluate the performance of
a decoder. It parametrically depends on $p$ and $q$, and we are interested
in its behavior for large systems $L\rightarrow \infty$. A value of $P_D=1$
indicates that the decoder $D$ correctly decodes every trajectory and
therefore allows us to restore the encoded amplitudes reliably. For high
error rates and/or few stabilizer measurements, $P_D$ is expected to drop to
$\tfrac{1}{2}$ because either the decoder fails to decode the system (viz.\
it tosses a coin to decide on a correction string) or the system is in the
disentangling phase and the amplitudes are lost to the environment.

In addition to the decoding probability $P_D$, we will use the \emph{mean
time to first failure (MTFF)} $T_D$ to compare different decoders. To
define $T_D$, consider a (ideally, infinitely long) trajectory $\mathcal{T}$
and for $t=0,1,\dots$ define $\mathcal{T}|_t$ as the first $t$ time steps of
$\mathcal{T}$ terminated by a full stabilizer measurement. We can then apply a
decoder $D$ to $\mathcal{T}|_t$ and compute $f^\sub{qm}_D(\mathcal{T}|_t)$.
Let $t^\sub{qm}_D(\mathcal{T})$ denote the first time step $t$ where
$f^\sub{qm}_D(\mathcal{T}|_t)<1$ and define the MTFF as the sample average
of this quantity:
\begin{align}
    T_D(p,q;L):=\langle\langle t^\sub{qm}_D\rangle\rangle\,.
\end{align}
We will refer to decoders as \emph{succeeding} for a given set of parameters
$p$ and $q$ if the MTFF grows superpolynomially in the limit $L\to\infty$;
by contrast, the MTFF of a \emph{failing} decoder grows sublinearly (typically
logarithmically) in this limit.

\subsection{Efficient numerical simulations}
\label{subsec:simulation}

To evaluate $P_D$ numerically by sampling, for each sample two steps are
required: First, a trajectory $\mathcal{T}$ of the PTIM must be generated,
and, second, based on the syndromes $S$, the decoder must be evaluated to
compute $f^\sub{qm}(\mathcal{T},C=D(S))$. The second step depends on the
decoding algorithm used and will be commented on in the respective sections
where we study the performance of different decoders. The sampling of PTIM
trajectories is described in the following.

Since both $E_i$ and $S_e$ belong to the Pauli group on $L$
qubits, and $\ket{\Psi_0}=\ket{\vec 0}$ is the unique state
stabilized by $\{\sigma_i^z\}$, the PTIM evolution can be
simulated exactly within the stabilizer formalism with only polynomial
overhead~\cite{Gottesman_1996,Gottesman_1998_Theory,Gottesman_1998,Aaronson_2004},
as stated by the Gottesmann-Knill theorem~\cite{Nielsen_2010}.

However, there is a more efficient method by mapping the trajectory to a
\emph{classical} stochastic process derived from the PTIM that operates on
a chain of $L$ classical \emph{bits} initialized in the configuration
$\vec m_0=\vec 0$ (we drop the ``classical'' in the following). Instead
of performing error measurements $E_i$ with probability $p$, we flip every
bit in every time step with probability $p/2$. This results in a bit flip
pattern $\flpat$ on the space-time lattice which determines the
final bit configuration $\vec m_T$ after $T$ time steps. We then perform
stabilizer measurements on the classical system just as we would on the PTIM
(only there is no collapse due to the measurement so we can do this
en bloc after generating the bit flip pattern); this yields a pattern $S^p$
and the corresponding measurement results $S^r$. Given the data $S=(S^p,S^r)$
sampled in this fashion, we can apply a given decoder $D$ to compute the
correction string $C=D(S)\in\{0,1\}^L$.
With the (trivial) evaluation function
\begin{equation}
	f^\sub{bi}(\flpat,C):=
    \begin{cases}
		1,\;& \text{$C=\vec m_T$}\\
		0,\;& \text{otherwise},
	\end{cases}
\end{equation}
we can evaluate every decoder also on the classical system. [The superscript
$^\sub{bi}$ stands for ``bits'' to distinguish this evaluation function from
its quantum counterpart defined in \cref{eq:fqm}].

The crucial point, which we prove in \cref{app:simulateClassical}, is the
relation
\begin{equation}
    \langle\langle f^\sub{qm}_D\rangle\rangle_\sub{qm}
    =\langle\langle f^\sub{bi}_D\rangle\rangle_\sub{bi},
    \label{eq:fqmfbi}
\end{equation}
which holds for all decoders $D$ (the subscripts to the sample averages
indicate sampling with respect to the full quantum model and the classical
analog described above, respectively). This allows us to sample the classical
process described above to evaluate the performance of decoders instead of
performing the full quantum mechanical simulation of the PTIM. This approach
results in a considerable speedup compared to the stabilizer formalism,
which enables us to study larger systems and/or larger samples to reduce
statistical fluctuations. We cross-checked the validity of \cref{eq:fqmfbi}
numerically on smaller systems using a full-fledged stabilizer simulation.

\section{Decoding with Majority Voting}
\label{sec:MVD}

\begin{figure*}[tbp]
	\includegraphics[width=0.8\linewidth]{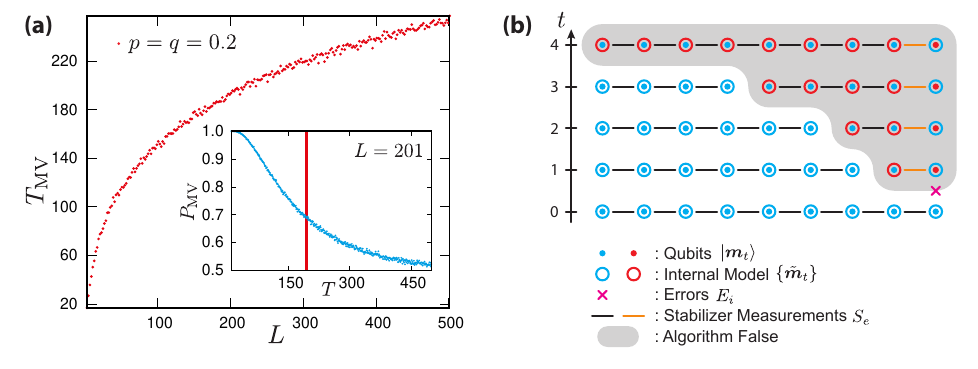}
	\caption{%
        \CaptionMark{Majority voting decoder (MVD).} 
        \lbl{a}~Mean time to first failure (MTFF) $T_\sub{MV}$ for the
        MVD for $p=q=0.2$ averaged over $5000$ trajectories as a function
        of the chain length $L$ (odd $L$). Inset: Decoding probability
        $P_\sub{MV}(p,q;L,T)$ for the same $p$ and $q$ and $L=201$ averaged
        over $10\,000$ trajectories as a function of the decoding time $T$. The
        vertical red line indicates the MTFF $T_\sub{MV}(p,q;L)$ for these
        parameters. These results demonstrate that the MTFF does \emph{not}
        grow exponentially with the system size. We checked that these results
        are representative and do not change for other parameters $p$ and $q$.
        \lbl{b}~Example trajectory that illustrates a crucial weakness
        of the MVD. The tentative bit pattern $\{\tilde{\vec m}_t\}$
        constructed by the decoder is indicated by circles; the true
        state of the system $\{\vec m_t\}$ by bullets. Note that this
        is an example where the state $\ket{\Psi(t)}$ has product form
        $\ket{\vec m_t}$ for all times $t$. The decoding result is $\hat
        C\ket{\Psi(T=4)}=\ket{\vec 1}\neq\ket{\vec 0}=\ket{\Psi_0}$, so
        $f^\sub{qm}(\mathcal{T},C=D(S))=0$.
    }
	\label{fig:MV}
\end{figure*}

\subsection{Algorithm}

We will now introduce and discuss our first decoder $D$. This decoder will
have the threshold $p=0$ except for the special point  $q=0$, and therefore
will fail to decode the stored quantum information for the PTIM. It serves
as preparation and motivation for the more sophisticated decoders in
\cref{sec:MWPM,sec:MLD}.

Let us first focus on the special case with $q=0$, where at every time
step \emph{all} stabilizers are measured and projective errors occur with
probability $p\in (0,1)$. This is the usual situation of quantum error
correction. Note, however, that our \emph{projective} error measurements
with probability $p$ effectively result in qubit flips with probability
$p/2$. This is in contrast to conventional treatments where errors are
modeled by \emph{unitary} operators, in which case the probability for an
error to occur and for a qubit to flip are identical.

If we exclude the exponentially unlikely situation where in a single
time step an error occurs on every qubit, it is straightforward to
check that, starting from $\ket{\Psi_0}=\ket{\vec 0}$, every round of
stabilizer measurements projects the system into a product state of
the form $\ket{\Psi(t)}=\ket{\vec m_t}$. In this situation, the PTIM
evolution becomes basically classical and guessing the qubit flips from
the stabilizer measurements $S^r$ is equivalent to decoding the classical
repetition code: To compute the final correction string $C$ that recovers
the initial state, $\hat C\ket{\Psi(T)}=\ket{C\oplus\vec m_T}=\ket{\vec
0}$, we split $C=\oplus_{t=1}^T C_{t}$ into correction strings $C_{t}$
that correct the qubit flips that occurred in time step $t$. To compute
$C_{t}$ from the accumulated syndrome data $S^r=(S^r_t)_{t=0,\dots,T}$, we
compare the syndrome measurements $S^r_{t-1}$ at time $t-1$ with the ones
$S^r_{t}$ in the subsequent time step. Since the syndromes are complete (all
stabilizers were measured, $q=0$), this allows for only two consistent flip
patterns $C_t\in\{C_{t,1},C_{t,2}\}$ with $C_{t,2}=\overline{C_{t,1}}$ (the
bar denotes the complementary bit string). The premise of the \emph{majority
voting decoder (MVD)} is to choose the one with \emph{fewer} flips, as this
is the more likely one for $p/2<1/2$. This choice is unique for chains of
odd length $L$; for consistency, we will stick to odd $L$ throughout the paper.
It is well-known that this decoder is perfect, i.e., it succeeds for
all $p/2<1/2$ almost surely in the limit $L\to\infty$, which is therefore
also true for the PTIM evolution at $q=0$. Note that in contrast to
conventional error models, where $p$ denotes the probability for (unitary)
qubit flips, the decoding transition in the PTIM appears at $p=1$, i.e.,
when \emph{all} qubits undergo an error measurement within each time step.
Our goal is now to generalize the MVD decoder to the case where some
stabilizers fail to be measured with probability $q>0$.

The basic procedure remains unchanged, i.e., we decompose $C=\oplus_{t=1}^T
C_{t}$ into corrections per time step and try to derive $C_t$ from the syndrome
measurements. Because of $q>0$, typically there will be gaps in the syndromes
$S^r_t$ where no measurement was performed. This lack of knowledge enlarges the
set of consistent flip patterns $C_t\in\{C_{t,1},C_{t,2},C_{t,3},\dots\}$. To
construct this list efficiently, we define a ``tentative'' qubit configuration
$\tilde{\vec{m}}_{t-1}\equiv \oplus_{\tau=1}^{t-1}C_\tau\oplus\vec 0$. By
construction, $\tilde{\vec{m}}_{t-1}$ is consistent with the (partial) syndrome
$S^r_{t-1}$. To construct $C_t$, we list all consistent flip patterns $C_{t,k}$
such that $\tilde{\vec{m}}_{t,k}\equiv C_{t,k}\oplus\tilde{\vec{m}}_{t-1}$
is consistent with the partial syndromes $S^r_t$. Following the rationale
of majority voting, we then choose for $C_t$ the $C_{t,k}$ with the fewest
flips. This defines our decoding algorithm $D:S\mapsto C=D(S)$, which we will
analyze in the next subsection.

\subsection{Results}

To assess the performance of majority voting, we computed the MTFF
$T_\sub{MV}$ as a function of the system size $L$ for fixed
parameters $p$ and $q$. The results are shown in \cref{fig:MV}~(a) for
the representative parameters $p=0.2=q$ and $L$ up to $501$. These results
demonstrate that the MTFF does \emph{not} scale superpolynomially with the
system size $L$, i.e., the decoding algorithm does not decode the system
efficiently. We checked that varying the parameters $p$ and $q$ does not
alter the result qualitatively, i.e., there seems to be no parameter regime
with $q>0$ where $T_\sub{MV}$ grows superpolynomially with $L$.

This behavior can be made plausible with a single trajectory that highlights
a crucial weakness of the MVD, see \cref{fig:MV}~(b). The
sketched trajectory has only a single error measurement $E_{9}$ between $t=0$
and $t=1$; furthermore, there is a single missing stabilizer measurement at
$t=1$ that creates a two-qubit segment connected by a syndrome measurement
$S_e=-1$. Consequently, the decoder can only toss a coin to decide on the
location of the error. In $50\,\%$ of the cases, its choice is incorrect, 
so the internal model $\tilde{\vec m}_{1}$ deviates from the true state
$\ket{\Psi(t=1)}=\ket{\vec m_1}$ on this segment. In the next steps, no errors
occur but there is a missing stabilizer in each time step. Because the majority
voting is restricted to segments of contiguous stabilizer measurements, the
decoder is forced to enlarge the discrepancy between $\tilde{\vec m}_t$ and
$\vec m_t$ until at $t=4$ the internal model is wrong everywhere. In total,
the decoder has to assume eight flips for its internal model $\{\tilde{\vec
m}_t\}$ whereas in reality a single error occurred. Because of its time-local
mode of operation to determine $C_t$, the MVD has no chance to find the true
error pattern deterministically.

There is also a more abstract perspective on this. Remember that the decoder
\emph{does work} for $q=0$, i.e., for the conventional repetition code. For
this, it is crucial that majority voting is applied to an extensive set of
qubits (namely, $L$). For $q>0$, the probability to find a segment of $l$
contiguous stabilizer measurements in a single time step is $q^2(1-q)^l$;
the average length of such a cluster is therefore $\overline l=(1-q)/q$. The
MVD, as defined above, performs majority voting on each of these segments
separately---it cannot keep track of correlations between the segments. Because
$\overline l$ is independent of $L$ and finite for $q>0$, these $\overline
l$-local decisions do not improve for $L\to\infty$.

Combining these findings, the crucial flaw of the MVD seems to be that it
composes the final correction string $C$ out of stepwise corrections $C_t$
that are the result of a time-local minimization procedure. Especially,
the decoder does not take into account the full syndrome data $S$
\emph{globally} but slices it into independent pieces; it operates, in a sense,
one-dimensional and ignores the two-dimensional space-time geometry
that comes with $S$. Thus, it cannot exploit correlations between disjoint
segments of contiguous syndrome measurements at any given time step. We will
overcome this issue with the much more sophisticated MWPM decoder in the next section.

\section{Decoding with Minimum Weight Perfect Matching}
\label{sec:MWPM}

\subsection{Algorithm}

\begin{figure*}[tbp]
	\includegraphics[width=\textwidth]{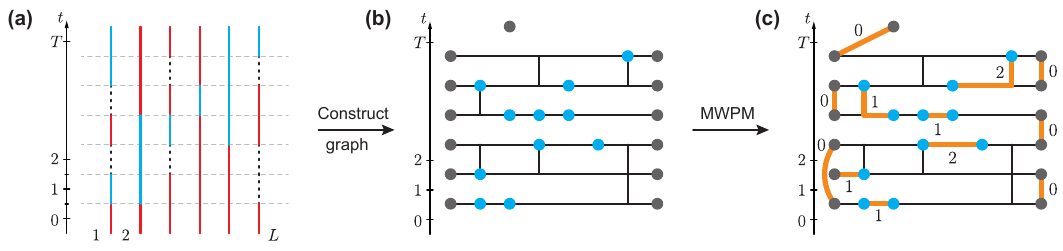}
	\caption{%
        \CaptionMark{Minimum weight perfect matching (MWPM) decoder---algorithm.} 
        \lbl{a}~Syndrome data of the trajectory in \cref{fig:system}~(b) on
        the dual space-time lattice; this is the input of the decoder. [The
        panel is mostly identical to \cref{fig:system}~(c); we reproduce it
        here for convenience.]
        \lbl{b}~Construction of the reduced, dual space-time lattice
        (graph) as the input for the MWPM algorithm. It consists of
        \emph{all} horizontal edges but only the \emph{dashed} vertical
        edges of the dual space-time lattice in \lbl{a}. The endpoints of
        blue line segments in \lbl{a} are labeled by blue nodes. To allow
        matchings to the edges, we add dummy nodes (gray) on the boundary,
        and an additional dummy node (top) to make the total number of
        nodes even. The dummy nodes are fully connected by dummy edges
        (not shown). All edges are weighted by their number of horizontal
        segments of the dual space-time lattice (= qubits). This weighted
        graph allows for a perfect matching of all blue and gray nodes by
        construction and is the input to the MWPM algorithm.
        \lbl{c}~Example of a perfect matching (orange lines). The weight of
        each matching path is given by the number of traversed horizontal
        edges (which correspond to assumed error measurements). The shown
        perfect matching is a MWPM as it
        minimizes the total weight (= number of assumed errors).
    }
	\label{fig:MWPMa}
\end{figure*}

The MWPM decoder takes into account the
full two-dimensional space-time geometry and derives from the syndromes
$S$ a possible error pattern $E^p$ with a \emph{globally} minimal number of
errors. This error pattern is then used to decide on one of the two decoding
strings $\{C,\overline{C}\}$. 
Our approach is motivated by the use of MWPM
for the decoding of two-dimensional surface codes
\cite{Duclos-Cianci_2010,Fowler_2012}, which is conceptually similar to the
decoding of a noisy, one-dimensional quantum repetition code \cite{Dennis_2002}
in that nodes on a two-dimensional lattice must be matched pairwise.


The decoder $D$ is defined by three steps: First, the error syndrome is
used to construct an abstract graph on which then, in the second step,
MWPM is performed. Finally, the found MWPM
is used to select one of the two decoding strings
$\{C,\overline{C}\}$. We now describe these three steps in detail, following
the illustrations in \cref{fig:MWPMa}; a motivation for this algorithm is
given below.

\begin{enumerate}
    \item In the first step, we construct from the space-time pattern of
    syndromes~$S$ in \cref{fig:MWPMa}~(a) the reduced dual space-time lattice
    (``graph'') in \cref{fig:MWPMa}~(b), augmented by dummy nodes and dummy
    edges. To this end, we keep \emph{all} horizontal edges of the lattice
    [dashed gray in (a)], but \emph{only} the vertical edges without syndrome
    measurement [dashed black in (a)]. We then highlight all nodes of the graph
    in (b) where a blue vertical line with $S_i=-1$ in (a) terminates. To
    allow for matchings to the boundaries, we add gray dummy nodes on the
    endpoints of horizontal edges. If the total number of highlighted nodes
    (blue and gray) is odd, we add an additional dummy node to the graph. We
    then connect all dummy nodes pairwise by dummy edges [for the sake
    of clarity, we do not show these in \cref{fig:MWPMa}~(b)]. Finally,
    we assign integer weights to all edges of the graph which count the
    number of horizontal segments (= qubits) traversed by these edges. In
    particular, all vertical and all dummy edges have weight zero.
    \item In the second step, we compute a MWPM
    of the blue and gray nodes on this graph. 
    A \emph{perfect} matching is given by pairwise connections of \emph{all}
    highlighted nodes along edges of the graph. A perfect matching has
    \emph{minimum weight} if the total sum of weights of all used edges
    is minimal.
    Minimum weight perfect matchings can be efficiently computed using
    the \emph{Blossom algorithm}~\cite{Edmonds_1965}; here we use an
    optimized implementation by Kolmogorov, known as \texttt{Blossom
    V}~\cite{Kolmogorov_2009}. The result is a MWPM, illustrated by orange
    paths in \cref{fig:MWPMa}~(c), where every horizontal edge traversed
    contributes $1$ to the total weight. The horizontal edges of the MWPM
    are the positions where the decoder assumes that a qubit was flipped.
    \item In the third and final step, we use the MWPM to decide on the
    correction string $C=(c_1,\dots,c_L)=D(S)$. We set $c_i=0$ if the qubit
    at position $i$ crosses an \emph{even} number of horizontal edges in the
    MWPM, and $c_i=1$ if it crosses an \emph{odd} number. In other words:
    the decoder assigns $c_i=0$ ($c_i=1$) to qubit $i$ if it assumes an even
    (odd) number of flips on this qubit.
\end{enumerate}

The motivation for this algorithm is that a perfect matching on the constructed
graph describes a \emph{possible} error pattern $E^p$ (if one ignores the
matchings on dummy edges between dummy nodes). This is so because the
paths of the perfect matching in combination with the blue vertical edges
in \cref{fig:MWPMa}~(a) (where syndrome measurements $S_i=-1$ signal flipped
adjacent qubits) form the closed boundaries of space-time regions where the
qubits might have been flipped. A given perfect matching therefore presupposes
projective errors on all its horizontal segments of the space-time lattice. In
particular, the total weight of the matching (= sum of all horizontal segments)
corresponds to the number of required errors. Computing an error pattern $E^p$
that (1) explains the observed syndrome $S$ and (2) minimizes the total number
of errors
is therefore equivalent to finding a MWPM on the
reduced space-time lattice. The matchings to dummy nodes on boundaries are
necessary because the endpoint of error strings that terminate on boundaries
is not signaled by a syndrome measurement.

Note that this approach is very similar to the decoding of topological quantum
memories like the surface code~\cite{Kitaev_2003,Bravyi_1998,Dennis_2002},
where decoding boils down to matching (= fusing) pairs of anyonic
excitations with a minimal amount of (unitary) errors. It is well-known
that MWPM is an efficient method to achieve
this~\cite{Duclos-Cianci_2010,Fowler_2012}. Indeed, the situation here can
be interpreted as the decoding problem of an anisotropic version of the
surface code with peculiar boundary conditions~\cite{Dennis_2002}.

\subsection{Results}

\begin{figure*}[tbp]
	\includegraphics[width=\textwidth]{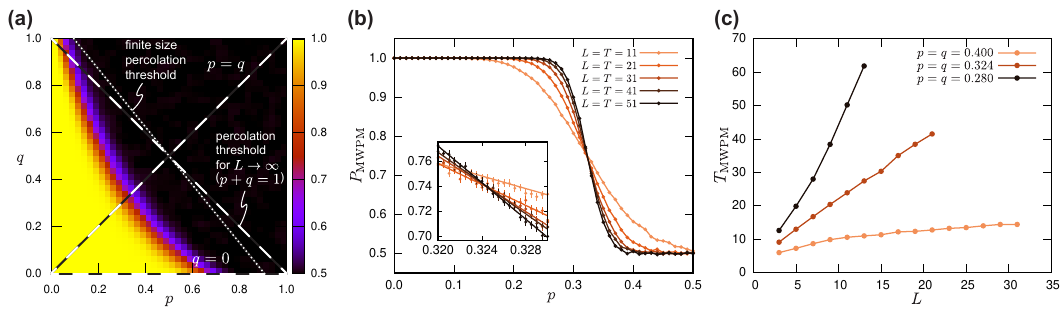}
	\caption{%
        \CaptionMark{Minimum weight perfect matching (MWPM) decoder---results.} 
        \lbl{a}~Decoding probability $P_{\sub{MWPM}}$ of the MWPM decoder as
        a function of $p,q\in[0,1]$ for $L=T=51$, computed from a sample of
        $20\,000$ trajectories for every datapoint. For comparison, we reproduce the percolation
        thresholds from \cref{fig:cluster} (dotted: numerical for finite-size
        system; dashed: exact for $L\to\infty$); they indicate the entanglement
        transition of the PTIM. The decoding phase is a proper subset of
        the entangling phase of the PTIM.
        \lbl{b}~Decoding probability $P_{\sub{MWPM}}$ along the
        diagonal $p=q$ in \lbl{a} for increasing system sizes
        $T=L=11,\dots,51$ and sampled over $50\,000$ trajectories for every datapoint. We
        find a crossing with negligible finite-size shift that gets
        sharper for $L=T\to\infty$ (inset), indicating an error threshold
        $p_\sub{thr}^\sub{MWPM}=q_\sub{thr}^\sub{MWPM}\approx 0.324$ away
        from the entanglement transition at  $p_c=q_c\approx 0.5$. The error
        bars in the inset represent the standard deviation of the samples.
        \lbl{c}~Mean time to first failure (MTFF) $T_\sub{MWPM}$ as a
        function of system size $L$ in the decoding phase ($p_1=0.280$), at
        the error threshold ($p_2=0.324$), and outside the decoding phase
        but in the entangling phase ($p_3=0.400$), sampled over $5\,000$
        trajectories for every datapoint. The MTFF grows exponentially with the system size in
        the decoding phase [cf.~\cref{fig:MV}~(a)].
    }
	\label{fig:MWPMb}
\end{figure*}

We start our analysis by computing the decoding probability $P_{\sub{MWPM}}$
for $L=T=51$ as a function of $p$ and $q$, \cref{fig:MWPMb}~(a). Note that it
is reasonable to scale $T\sim L$ when studying the MWPM decoder; for simplicity,
we set $T=L$ in the following. The results demonstrate that there is a finite
region in parameter space where decoding succeeds. As expected, this region
is fully contained in the regime where the PTIM is in the entangling phase
(cf.~\cref{fig:cluster}); it is considerably smaller though. There is a clear
transition between the region where decoding succeeds ($P_{\sub{MWPM}}\approx
1$) and the region where it fails ($P_{\sub{MWPM}}\approx 0.5$). We checked
that this transition becomes sharper for larger systems (it also shifts
slightly; we discuss this in more detail below).
Let us comment on a few peculiarities. First, on the $q$ axis ($p=0$)
the decoder always succeeds because there are no errors to be corrected.
This intuition is confirmed by our results. On the $p$ axis ($q=0$), one would
expect the decoder to be successful as well because MWPM is equivalent to
global majority voting for $q=0$, i.e., when every stabilizer is measured (and
we know that the MVD succeeds in this special case for all $p/2<1/2$). However,
our results suggest that the decoder \emph{fails} for $p\gtrsim 0.7$. We
confirmed that this is a (strong) finite-size effect which can even be
described analytically (see \cref{app:finitesize}). For $L=T\to\infty$,
the decoding region will indeed grow on the $p$ axis until it reaches $p=1.0$.

To assess the finite-size scaling on the diagonal, we computed
$P_{\sub{MWPM}}$ as a function of $p=q$ for increasing system sizes $L=T$,
see \cref{fig:MWPMb}~(b). We find a clear crossing with almost negligible
finite-size shift (in contrast to $q=0$) that becomes sharper in the limit
$L=T\to\infty$. We conclude that the decoding phase is indeed a proper
subset of the entangling phase and pinpoint the decoding threshold at
$p_\sub{thr}^\sub{MWPM}=q_\sub{thr}^\sub{MWPM}\approx 0.324$. Note that on
the diagonal the entanglement transition takes place at $p_c=q_c=0.5$. In
the intermediary regime $p_\sub{thr}<p=q<p_c$ we face the peculiar situation
that the encoded amplitudes of the logical qubit survive the monitoring
by the environment but are inaccessible using only the syndrome data $S$
and MWPM decoding.



Now that we know the phase diagram of the MWPM decoder, we can also evaluate
the MTFF $T_\sub{MWPM}$. The decoding transition is
also visible in this quantity. To demonstrate this, we pick three values for
$p$ on the diagonal $p=q$: $p_1=0.280$ in the decoding phase, $p_2=0.324$ on
the phase boundary, and $p_3=0.400$ outside the decoding phase (but in the
entangling phase). In \cref{fig:MWPMb}~(c), we plot the MTFF $T_\sub{MWPM}$
as a function of system size $L=T$. It shows the expected exponential behavior
in the decoding phase, while it seems to grow linearly ($T_\sub{MWPM}\sim L$)
at criticality and sub-algebraically [$T_\sub{MWPM}\sim \log(L)$] away from
the decoding phase.

\section{Maximum Likelihood Decoding}
\label{sec:MLD}

So far, we have demonstrated the existence of a nontrivial decoding threshold
for our model, i.e., a transition at a critical error rate up
to which it is possible to retrieve the stored quantum information knowing
only the syndrome $S$. While the entanglement transition provides an upper
bound on this critical error rate, the MWPM decoder presented in the previous section
provides a lower bound on this threshold.
Decoders that achieve this threshold are MLDs. These decoders are
defined as the ones that, given the syndrome data $S$, choose the correction
string that is most likely correct. This is done by calculating probabilities
for classes of trajectories that yield the same syndrome data $S$ and final
state $\hat{C}\ket{\Psi_0}$. We define the probability of such a class as
\begin{equation}
	P^\sub{qm}_{f^\sub{qm}}(C |S)
    :=\sum_{\mathcal{T}|_S}P^\sub{qm}(\mathcal{T})\cdot f^\sub{qm}(\mathcal{T},C),
    \label{eq:MLDgeneral}
\end{equation}
where $\mathcal{T}|_S$ restricts the summation to trajectories with the
syndrome $S$. The output of the MLD is the correction
string $C$ that maximizes \cref{eq:MLDgeneral} for a given syndrome
$S$. Clearly, the maximum likelihood decoder provides the highest possible
decoding probability
\begin{equation}
    P_{\rm \scriptscriptstyle MLD}(p,q;L,T)=\sup_D P_D(p,q;L,T),
\end{equation}
and therefore is the best possible decoder. While in many cases it is
impossible (or unknown how) to implement a maximum likelihood decoder
efficiently for a given error correction code, below we demonstrate the
efficient implementation of such a decoder for the PTIM.

Performing the sum in \cref{eq:MLDgeneral} is highly nontrivial because
it is constrained by the syndrome and typically contains an exponentially
large number of terms. Recently, an efficient implementation of an
MLD for the (perfect) surface code has been demonstrated by
Bravyi \emph{et al.}~\cite{Bravyi_2014} using a clever resummation technique that is based
on an equivalent formulation in terms of a quadratic fermion theory.
Here we implement the maximum likelihood decoding for our system in two steps:
\begin{enumerate}
    \item We introduce a classical stochastic process with random bit flips
        and stabilizer measurements, and prove that defining the MLD based
        on the decoding probability $P^\sub{bi}_{f^\sub{bi}}(C|S)$ for
        this classical model is equivalent to implementing the MLD of its
        quantum mechanical pedant.
    \item The decoding probability $P^\sub{bi}_{f^\sub{bi}}(C |S)$  for this
        classical model is then efficiently evaluated
        using the resummation techniques developed by
        Bravyi \emph{et al.}~\cite{Bravyi_2014}.
\end{enumerate}

\begin{figure*}[tbp]
	\centering
	\includegraphics[width=0.9\textwidth]{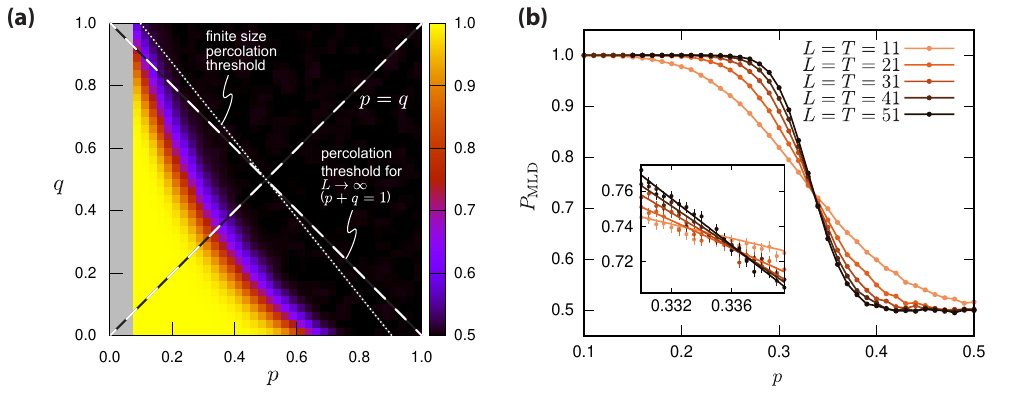}
	\caption{%
		\CaptionMark{Maximum likelihood decoder (MLD).}
		\lbl{a}~Decoding probability $P_{\sub{ML}}$ of the ML decoder as a
		function of $p,q\in[0,1]$ for $L=T=41$, computed from a sample of
		$20\,000$ trajectories for every datapoint. For comparison, we reproduce the percolation
		thresholds from \cref{fig:cluster} (dotted: numerical for finite-size
		system; dashed: exact for $L\to\infty$); they indicate the entanglement
		transition of the PTIM. As for the MWPM decoder, the decoding phase
		is a proper subset of the entangling phase of the PTIM. We omit data
		for $p\leq 0.05$ because in this regime the MLD algorithm becomes
		numerically unstable and the results are not trustworthy.
		\lbl{b}~Decoding probability $P_{\sub{MWPM}}$ along the diagonal
		$p=q$ in \lbl{a} for increasing system sizes $T=L=11,\dots,51$
		and sampled over $50\,000$ trajectories for every datapoint. Again we find a
		crossing with negligible finite-size shift that gets sharper
		for $L=T\to\infty$ (inset), indicating an error threshold
		$p_\sub{thr}^\sub{ML}=q_\sub{thr}^\sub{ML}\approx 0.336$ away from
		the entanglement transition at $p_c=q_c\approx 0.5$. The error bars
		in the inset represent the standard deviation of the samples.
	}
	\label{fig:MLD}
\end{figure*}

The classical stochastic process derived from the PTIM  has already been introduced
in \cref{subsec:simulation}, and is reviewed here again. It describes
a chain of $L$ classical \emph{bits} with all bits initialized in state
$0$. The dynamics is governed by random bit flips with probability $p/2$,
resulting in a bit flip pattern $\flpat$ on the space-time lattice. In every
time step, we flip the bits according to $\flpat$ and then perform stabilizer
measurements with probability $1-q$, which yields the stabilizer pattern
$S^p$. The input of the decoder is the combination of the stabilizer pattern
and the corresponding results $S=(S^p,S^r)$, just as for the quantum system.

The crucial point is that there is a straightforward way to adapt Vargo's
algorithm to the classical variant of our system. We then use this algorithm
to compute the total probability $P^\text{bi}_{f^\text{bi}}(C|S)$ of a fixed
correction string $C$ by summing over all consistent bit patterns. This
results in an algorithm with a runtime that scales polynomially [like
$O(L^4)$] with system size $L=T$, and depends parametrically
on the error probability $p$ and syndrome failure rate $q$.
The decoder $D^\text{bi}_\text{ML}(S)$ then returns the correction string
that maximizes $P^\text{bi}_{f^\text{bi}}(C|S)$.

At this point, we have a working MLD for the
classical system. Due to the equivalence between the classical and the
quantum trajectories [recall \cref{eq:fqmfbi}], one can prove rigorously
that this algorithm satisfies the conditions for a
MLD also for the projective error model, i.e., $\langle\langle
f^\sub{qm}_{D^\sub{qm}_\sub{ML}}\rangle\rangle_\text{qm}=\langle\langle
f^\sub{qm}_{D^\sub{bi}_\sub{ML}}\rangle\rangle_\text{qm}$. This means
that the MLD of the classical system $D^\sub{bi}_\sub{ML}$ performs
just as well as the MLD of the quantum model $D^\sub{qm}_\sub{ML}$ when
applied on quantum trajectories; hence we will just refer to the classical
decoder $D^\sub{qm}_\sub{ML}$ as MLD in the
following. As the proof is rather lengthy and quite technical, we defer it
to \cref{app:MLDProof}.

\subsection{Results}

As for the MWPM decoder, we compute the decoding probability $P_\sub{ML}$
for the ML decoder as a function of $p,q\in[0,1]$ for a square space-time
lattice $L=T=41$, see \cref{fig:MLD}~(a). Note that the ML algorithm is
computationally more expensive than the MWPM algorithm, which is why we
stick to smaller systems for reliable statistics.
The results are very similar to MWPM decoding. Most importantly, the
decoding phase still seems to be a proper subset of the PTIM entangling
phase. Contrary to the MWPM decoder, we encounter numerical instabilities
for very low error rates $p\lesssim 0.05$ where the results become erratic
and seemingly random. This instability is a consequence of the very small
probabilities of specific error patterns and has already been noted in
Ref.~\cite{Bravyi_2014}. We evade this technical issue by omitting the
unreliable data in \cref{fig:MLD}~(a), as we are mainly interested in the
phase boundary anyway.

To check for finite-size effects, we plot $P_\sub{ML}$ on the diagonal $p=q$
as a function of the system size $L=T=11,\dots,51$, \cref{fig:MLD}~(b). As
for the MWPM decoder, there is a clear crossing with only a small finite-size
shift (inset); the transition gets sharper for larger systems and indicates
an error threshold $p_\sub{thr}^\sub{ML}=q_\sub{thr}^\sub{ML}\approx 0.336$,
slightly larger than the MWPM threshold $p_\sub{thr}^\sub{MWPM}\approx 0.324$.
This difference is small but nonetheless shows that the ML decoder performs
slightly better than the MWPM decoder, as expected.
However, these findings also show that the MWPM decoder is already a
near-optimal decoder and cannot be improved significantly.
The rather small improvement of ML over MWPM
decoding is in agreement with previous results for surface codes \cite{Bravyi_2014,Fowler_2012}.


\section{Summary}
\label{sec:summary}

In this paper, we studied the error correction capabilities of the
PTIM, a stochastic model of two competing
classes of projective measurements that is characterized by an entanglement
transition. We interpret the competing measurements as stabilizer measurements
of the quantum repetition code and error measurements by the environment,
respectively. In the entangling phase of the PTIM, the system acts as a quantum
memory that preserves the amplitudes of a single logical qubits from being
accessed by the environment. Our goal was to study methods for retrieving
these amplitudes without having access to the measurements by the environment.

In a first attempt, we generalized the MVD---which is
known to work for (quantum) repetition codes---to our setting. Numerical
results suggested that this approach fails for generic parameters of the
PTIM. We provided an intuition for this failure and used it as a starting
point to construct our second decoding algorithm based on
MWPM. This decoder makes use of the full syndrome data and
numerics revealed that it successfully retrieves the encoded amplitudes
for a nontrivial range of parameters which, however, does not exhaust the
complete entangling phase of the PTIM. This result suggested the existence
of a parameter regime where quantum amplitudes cannot be accessed by the
environment but, at the same time, remain inaccessible without having full
access to the state of the system. To assess this hypothesis rigorously,
we introduced a third decoder---the MLD---and showed
that it improves only slightly on MWPM. Because the
MLD is the optimal decoder for our system, we concluded
that there is indeed an intermediary regime where the encoded amplitudes are
neither accessible to the environment (through error measurement) nor to
the observer (through syndrome measurements). This result also shows that
it is impossible to pinpoint the entanglement transition by measuring the
syndrome data alone. However, from a practical point of view, the
MWPM decoder seems to be the better choice because it is
already a near-optimal and computationally less expensive than the MLD.

\begin{acknowledgements}
	This project has received funding from the German Federal Ministry of Education and Research (BMBF) under the grants QRydDemo and MUNIQC-Atoms.
\end{acknowledgements}


\bibliographystyle{bibstyle}
\bibliography{bibliography}


\clearpage
\appendix

\section{Classical simulation of the Quantum System}
\label{app:simulateClassical}

In the main text, we used the quantity $P_D\!\left(p,q;L,T\right)$ to
evaluate the performance of decoders. Here we intend to simulate realistic
quantum systems to test our decoders and evaluate the decoding probability
$P_D^\sub{qm}\!\left(p,q;L,T\right)=\langle\langle f^\sub{qm}_D\rangle\rangle$,
which is a functional of the decoder $D$. In the following, we will prove
that $\langle\langle f^\sub{qm}_D\rangle\rangle_\text{qm}=\langle\langle
f^\sub{bi}_D\rangle\rangle_\text{bi}$, where we also distinguish between the
sampling of a quantum and a classical system. This allows us to evaluate the
performance of a decoder $D$ on the quantum system by simulating classical
systems with bits.

Before we start the proof, we remind our readers that any quantum system
is defined by the pattern of error measurements $E^p$, the pattern of
stabilizer measurements $S^p$, and the corresponding measurement results $E^r$
and $S^r$. However, in our use case the results of the error measurements
are not important, so we will drop them in our calculations. On the other
hand, a classical trajectory of bits is fully determined by the pattern of
bit flips $\tilde{E}^p$ but there still exists a pattern of stabilizer
measurements $S^p$ and the corresponding results $S^r$. To facilitate
our calculations, we want to introduce the number of measurements in
a pattern which we denote as $\left|S^p\right|$, $\left|E^p\right|$,
and $\left|\tilde{E}^p\right|$. Furthermore, we introduce the concept of
reduced measurement patterns $E^p_\text{red}$ and $S^p_\text{red}$, which are
subsets of the original patterns but only include the measurements without
a predetermined measurement result. For example, measuring a stabilizer twice
in a row will not change the result, so the second measurement would not be
part of the reduced pattern.

For any system, the sample average is
\begin{equation}
    \langle\langle f^\alpha_D\rangle\rangle_\alpha
    =\sum_{\mathcal{T}^\alpha\in\left\{\mathcal{T}^\alpha\right\}}P^\alpha\!
    \left(\mathcal{T}^\alpha\right)\cdot f^\alpha\!\left(\mathcal{T}^\alpha;D\!\left(S\right)\right),
\end{equation}
where $\mathcal{T}^\alpha$ indicates a trajectory in the system $\alpha$
and $\left\{\mathcal{T}^\alpha\right\}$ is the set of all possible
trajectories. $P^\alpha\!\left(\mathcal{T}^\alpha\right)$ is the probability
of $\mathcal{T}^\alpha$ to occur in the system $\alpha$ and $f^\alpha$
denotes the evaluation function for $D$ in $\alpha$, where we make clear
that the decoder $D(S)$ is a function which depends only on the stabilizers
$S$ which are part of the trajectory.

Now focus on the quantum system. The probability of a single trajectory
$\mathcal{T}^\sub{qm}$ occurring is
\begin{equation}
	\begin{split}
		P^\sub{qm}\!\left(\mathcal{T}^\sub{qm}\right)
        :={}&\underbrace{p^{\left|E^p\right|}
        \left(1-p\right)^{LT-\left|E^p\right|}}_{P^\sub{qm}\!\left(E^p\right)}
        \\
		&\cdot\underbrace{q^{\left(L-1\right)T-\left|S^p\right|}
        \left(1-q\right)^{\left|S^p\right|}}_{P^\sub{qm}\!\left(S^p\right)}
        \\
		&\cdot\underbrace{\left(\frac{1}{2}\right)^{\left|S_\sub{red}^p\right|}}_{P^\sub{qm}\!\left(S^r\vert E^p,S^p\right)},
	\end{split}
\end{equation}
with the number of stabilizer and error measurements $\left|S^p\right|$
and $\left|E^p\right|$. We can insert this for the sample average:
\begin{widetext}
	\begin{align}
		\langle\langle f^\sub{qm}_D\rangle\rangle&
        =\sum_{\mathcal{T}^\sub{qm}\in\left\{\mathcal{T}^\sub{qm}\right\}}
        P^\sub{qm}\!\left(\mathcal{T}^\sub{qm}\right)\cdot f^\sub{qm}\!\left(\mathcal{T}^\sub{qm};D\!
        \left(S\right)\right)
        \\
		&=\sum_{E^p}P^\sub{qm}\!\left(E^p\right)\sum_{S^p}P^\sub{qm}\!
        \left(S^p\right)\sum_{S^r_{\vert E^p,S^p}}P^\sub{qm}\!\left(S^r\vert E^p,S^p\right)f^\sub{qm}\!
        \left(\mathcal{T}^\sub{qm};D\!\left(S\right)\right).
		\label{eq:sampleAverageQM}
	\end{align}
\end{widetext}
Here we sample the trajectories by first sampling over all error patterns,
then sampling over all stabilizer patterns, and lastly sampling over the
different possible results of the stabilizer measurements. (The results of
the error measurements have no effect on any of our observations and can
thus be ignored.)  We can do the same for a classical system of bits. Here
the probability of a classical trajectory $\mathcal{T}^\sub{bi}$ is
\begin{equation}
	\begin{split}
		P^\sub{bi}\!\left(\mathcal{T}^\sub{bi}\right)
        ={}&\underbrace{\left(\frac{p}{2}\right)^{\left|\flpat\right|}
        \left(1-\frac{p}{2}\right)^{LT-\left|\flpat\right|}}_{P^\sub{bi}\!\left(\flpat\right)}
        \\
		&\cdot\underbrace{q^{\left(L-1\right)T-\left|S^p\right|}
        \left(1-q\right)^{\left|S^p\right|}}_{P^\sub{bi}\!\left(S^p\right)}.
	\end{split}
	\label{eq:P(Tbi)}
\end{equation}
While it is possible to sample bit flip patterns $\flpat$ with the probability
$\nicefrac{p}{2}$ on every site in every time step, it will prove useful to
use a different approach. The idea is that bit flips can also be sampled
by first sampling a pattern $E^p$ of \emph{potential} bit flips using the
probability $p$, and successively sampling $\flpat$ from $E^p$ using the
probability $\nicefrac{1}{2}$. To do so, we add the pattern $E^p$ to the
trajectory $\mathcal{T}^\sub{bi}$ and define the following adjusted probability:
\begin{equation}
	\begin{split}
		P^\sub{bi}\!\left(\mathcal{T}^\sub{bi};E^p\right)
        :={}&\underbrace{p^{\left|E^p\right|}\left(1-p\right)^{LT-\left|E^p\right|}}_{P^\sub{bi}\left(E^p\right)}
        \\
		&\cdot\underbrace{q^{\left(L-1\right)T-\left|S^p\right|}
        \left(1-q\right)^{\left|S^p\right|}}_{P^\sub{bi}\left(S^p\right)}
        \\
		&\cdot\underbrace{\left(\frac{1}{2}\right)^{\left|E^p\right|}}_{P^\sub{bi}\left(\flpat\vert E^p\right)}.
	\end{split}
	\label{eq:P(Tbi;Ep)}
\end{equation}
This probability is related with \cref{eq:P(Tbi)} via
\begin{equation}
	P^\sub{bi}\!\left(\mathcal{T}^\sub{bi}\right)
    =\sum_{E^p_{\vert\mathcal{T}^\sub{bi}}}P^\sub{bi}\!
    \left(\mathcal{T}^\sub{bi};E^p\right).\label{eq:pseudoclassicalRelation}
\end{equation}
Here we take the sum over all patterns $E^p$ which are consistent with the
classical trajectory $\mathcal{T}^\sub{bi}$. This relation is proven in
\cref{sec:pseudoclassicalSampling}.

The new quantity defined in \cref{eq:P(Tbi;Ep)} allows us to write the
classical sampling average:
\begin{align}
    \langle\langle f^\sub{bi}_D\rangle\rangle_\text{bi}
    =&\sum_{\mathcal{T}^\sub{bi}}P^\sub{bi}\!\left(\mathcal{T}^\sub{bi}\right)
    \cdot f^\sub{bi}\!\left(\mathcal{T}^\sub{bi};D\!\left(S\right)\right)
    \\
    =&\sum_{S^p}\sum_{\flpat}\sum_{E^p_{\vert\mathcal{T}^\sub{bi}}}P^\sub{bi}\!
    \left(\mathcal{T}^\sub{bi};E^p\right)\cdot f^\sub{bi}\!\left(\mathcal{T}^\sub{bi};D\!\left(S\right)\right)
    \\
    \begin{split}
    =&\sum_{E^p}P^\sub{bi}\!\left(E^p\right)\sum_{S^p}P^\sub{bi}\!\left(S^p\right)
    \\
    &\cdot\sum_{\flpat\in E^p}P^\sub{bi}\!\left(\flpat\vert E^p\right)f^\sub{bi}\!
    \left(\mathcal{T}^\sub{bi};D\!\left(S\right)\right).
    \end{split}
    \label{eq:sampleAverageBi}
\end{align}
Here we used the fact that
$\sum_{\flpat}\sum_{E^p_{\vert\flpat}}\cdots=\sum_{E^p}\sum_{\flpat_{\vert
E^p}}\cdots$ and wrote
$\sum_{E^p_{\vert\mathcal{T^\sub{bi}}}}\cdots=\sum_{E^p_{\vert\flpat}}\cdots$
and $\sum_{\flpat_{\vert E^p}}\cdots=\sum_{\flpat\in E^p}\cdots$.

Comparing the quantum system in \cref{eq:sampleAverageQM} and the classical
system in \cref{eq:sampleAverageBi}, we see that the samplings only differ in
the last sum, where in the quantum case the sampling goes over the possible
stabilizer results and in the classical case the sampling goes over the
possible bit flips. In the following, we will show that the following equation:
\begin{equation}
	\begin{split}
		&\sum_{S^r_{\vert E^p,S^p}}P^\sub{qm}\!\left(S^r\vert E^p,S^p\right)f^\sub{qm}\!
        \left(\mathcal{T}^\sub{qm};D\!\left(S\right)\right)
        \\
		&=\sum_{\flpat\in E^p}P^\sub{bi}\!\left(\flpat\vert E^p\right)f^\sub{bi}\!
        \left(\mathcal{T}^\sub{bi};D\!\left(S\right)\right)
	\end{split}
\end{equation}
holds. As a first step, we insert the probabilities defined above:
\begin{equation}
	\begin{split}
		&\sum_{S^r_{\vert E^p,S^p}}\left(\frac{1}{2}\right)^{\left|S_\sub{red}^p\right|}f^\sub{qm}\!
        \left(\mathcal{T}^\sub{qm};D\!\left(S\right)\right)
        \\
		&=\sum_{\flpat\in E^p}\left(\frac{1}{2}\right)^{\left|E^p\right|}f^\sub{bi}\!
        \left(\mathcal{T}^\sub{bi};D\!\left(S\right)\right).
	\end{split}
\end{equation}
First, we focus on the classical right-hand side and make use of the fact that 
flipping a bit twice with a $\unit[50]{\%}$ probability yields the same sampling 
as flipping it once with the same probability. This means that we can sample
$\flpat$ just on the reduced error pattern $E_\sub{red}^p$ without changing
the result. On the quantum mechanical left-hand side of the equation, we can
see that the sampling over all possible stabilizer measurement results is
in reality just a sampling over the reduced stabilizer results. All other
results are determined by that. Thus, we find
\begin{equation}
	\begin{split}
		&\sum_{{S_\sub{red}^r}_{\vert E^p,S^p}}\left(\frac{1}{2}\right)^{\left|S_\sub{red}^p\right|}
        f^\sub{qm}\!\left(\mathcal{T}^\sub{qm};D\!\left(S\right)\right)
        \\
		&=\sum_{\flpat\in E_\sub{red}^p}\left(\frac{1}{2}\right)^{\left|E_\sub{red}^p\right|}
        f^\sub{bi}\!\left(\mathcal{T}^\sub{bi};D\!\left(S\right)\right).
	\end{split}
	\label{eq:proofEqual}
\end{equation}
We know that the patterns $E^p$ and $S^p$ determine whether or not the
original cluster survives in the quantum system. Therefore, we can consider
two cases now.

\begin{figure*}[tbp]
	\centering
	\includegraphics[width=\textwidth]{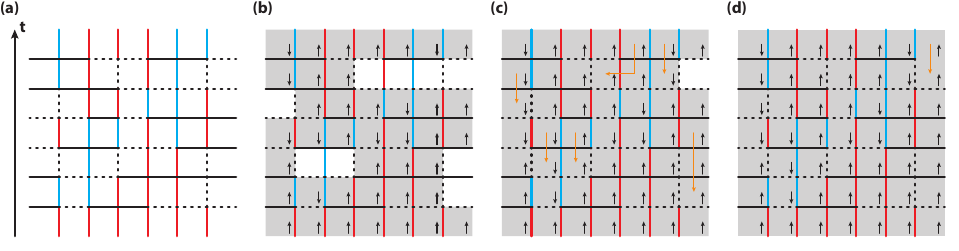}
	\caption{%
	\CaptionMark{Visual guide to identify terms in \cref{eq:proofEqual}
	if the original Bell cluster survives in the trajectory.}
    \lbl{a}~Here we show
	a grid $\mathcal{L}$ with the measurements of a quantum mechanical
	trajectory. The colors are defined as in \cref{fig:system}. 
    \lbl{b}~Wherever the original product state lives (gray plaquettes),
    squares can be associated with a spin orientation  (we indicate them with
	arrows).
    \lbl{c}~By backtracking the spin orientations  in time, more
	plaquettes of $\mathcal{L}$ can be associated with a spin orientation
	(which does not correspond to the quantum trajectory). \lbl{d} All
	remaining plaquettes can be backtracked by removing all $E^p\backslash
	E_\sub{red}^p$ from $\mathcal{L}$. The associated configuration of
	spins on $\mathcal{L}$ can be identified with a classical trajectory
	in \cref{eq:proofEqual}.
    } 
    \label{fig:QmBi}
\end{figure*}

\begin{enumerate}
	\item \emph{Cluster survives.} If the sampling
	of $E^p$ and $S^p$ is such that the original Bell
	cluster in the quantum mechanical system survives, it is
	$\left|S_\sub{red}^p\right|=\left|E_\sub{red}^p\right|$. Both sums
	have an equal number of terms and we can identify them with each
	other in pairs as we will explain now. First note that if the
	original cluster survives, the evaluation function $f^\sub{qm}$
	either returns value $1$ or $0$. Consider one term on the
	left side of \cref{eq:proofEqual}, which corresponds to a single
	quantum mechanical trajectory [see \cref{fig:QmBi}~(a)]. On the
	plaquettes of $\mathcal{L}$ on which the original cluster lives, we
	can easily determine the spin orientations as they are polarized in
	$z$ direction. This is the area marked gray in \cref{fig:QmBi}~(b). As
	we find the final state $\Ket{\psi_\text{f}}$ of the trajectory,
	this process already determines the value of the evaluation function
	$f^\sub{qm}\!\left(\mathcal{T}^\sub{qm};D\!\left(S\right)\right)$. Note
	that the trajectory on the gray plaquettes in \cref{fig:QmBi}~(b) has
	classical characteristics. To associate the quantum trajectory with
	a classical trajectory, the spin orientations on gray plaquettes
	can be artificially backtracked in time [without crossing error
	measurements, see \cref{fig:QmBi}~(c)]. Any missing plaquettes can be
	eliminated by deleting all trivial error measurements $E^p\backslash
	E_\sub{red}^p$ from $\mathcal{L}$. This allows us to backtrack
	those plaquettes too and assign every position in $\mathcal{L}$
	with a spin orientation in the $z$  direction (which is \textit{not}
	the real quantum trajectory). However, it is clear that this spin
	configuration can be considered as a valid trajectory on classical bits
	and be found as one term on the right side of \cref{eq:proofEqual},
	which obviously shares the same value of the evaluation function
	$f^\sub{qm}\!\left(\mathcal{T}^\sub{qm};D\!\left(S\right)\right)=f^\sub{bi}\!\left(\mathcal{T}^\sub{bi};D\!\left(S\right)\right)$.
	This procedure allows us to determine a corresponding classical
	trajectory for every quantum trajectory and thus pair up terms on
	both sides of \cref{eq:proofEqual} which share the same value. Thus,
	\cref{eq:proofEqual} holds.
	\item \emph{Cluster does not survive.} If the sampling of
	$E^p$ and $S^p$ is such that the original Bell cluster in
	the quantum mechanical system does not survive, we already
	know that [for any reasonable decoder $D\!\left(S\right)$]
	$f^\sub{qm}\!\left(\mathcal{T}^\sub{qm},D\!\left(S\right)\right)=\nicefrac{1}{2}$.
	On the other hand, it is
	$\left|S_\sub{red}^p\right|=\left|E_\sub{red}^p\right|-1$. We insert
	those relations into \cref{eq:proofEqual} and find
	\begin{equation}
		\begin{split}
			&\sum_{{S_\sub{red}^r}_{\vert E^p,S^p}}
            \left(\frac{1}{2}\right)^{\left|S_\sub{red}^p\right|}\cdot \frac{1}{2}
            \\
			&=\sum_{\flpat\in E_\sub{red}^p}\left(\frac{1}{2}\right)^{\left|S_\sub{red}^p+1\right|}
            f^\sub{bi}\!\left(\mathcal{T}^\sub{bi};D\!\left(S\right)\right).
		\end{split}
	\end{equation}
	However, there are twice as many classical trajectories than
	there are quantum trajectories and half of them are evaluated as
	$f^\sub{bi}\!\left(\mathcal{T}^\sub{bi};D\!\left(S\right)\right)=1$
	(for the other half of the trajectories, the evaluation function
	vanishes). Therefore, also in this case \cref{eq:proofEqual} holds.
\end{enumerate}
This proves, that $\langle\langle
f^\sub{qm}_D\rangle\rangle_\text{qm}=\langle\langle
f^\sub{bi}_D\rangle\rangle_\text{bi}$ independent on the decoder $D$. As
a consequence the performance of any decoder $D$ on the quantum system,
can be evaluated on classical systems.

\section{Proof of the Maximum Likelihood Decoder}
\label{app:MLDProof}

We will prove that the MLD introduced in
\cref{sec:MLD}, for which we assumed the system to be entirely classical
with bit flips instead of error measurements, deserves its name also in the
quantum mechanical system. Thus we will prove that the decoder yields the
most likely quantum mechanical spin configuration and therefore gives us
the best probability of decoding the system that can be achieved.

\subsection{Some Thoughts in Advance}

Consider two discrete variables $A$ and $B$, a function $P\!\left(B\right)\ge
0\forall B$ and a function $f\!\left(A;B\right)$. Then we can find an upper
bound to the following supremum:
\begin{equation}
	\begin{split}
		&\sup_A\left[\sum_BP\!\left(B\right)f\!\left(A;B\right)\right]
        \\
		\le&\sum_BP\!\left(B\right)\sup_A\left[f\!\left(A;B\right)\right].
	\end{split}
\end{equation}
Here we consider the sum over all $B$ and find the supremum out of all values
of $A$.

However if we now consider $A\!\left(B\right)$ to be a function out of the
set of all functions that take the values $B$ as input, we can actually
reach the upper bound:
\begin{equation}
	\begin{split}
		&\sup_A\left[\sum_BP\!\left(B\right)f\!\left(A\!\left(B\right);B\right)\right]
        \\
		=&\sum_BP\!\left(B\right)\sup_A\left[f\!\left(A\!\left(B\right);B\right)\right].
	\end{split}
	\label{eq:MLDProofMaximum}
\end{equation}
This is due to the fact that we can construct a function $A\!\left(B\right)$
that for every value of $B$ maximizes the function $f$ (no matter the form
of $f$). We will use this later.

\subsection{The Proof}

In general, we evaluate a decoder $D\!\left(S\right)$ in a system $\alpha$
via the function
\begin{equation}
	P^\alpha_D\!\left(p,q;L,T\right)=\langle\langle f^\alpha_D\rangle\rangle_\alpha,
\end{equation}
which is the average value of the evaluation function $f^\alpha$ for many
trajectories $\mathcal{T}^\alpha$. We can express this value by summing
over all possible trajectories $\left\{\mathcal{T}^\alpha\right\}$ with
the probability $P^\alpha\!\left(\mathcal{T}^\alpha\right)$ of a single
trajectory occurring:
\begin{equation}
	\langle\langle f^\alpha_D\rangle\rangle_\alpha
    =\sum_{\mathcal{T}^\alpha\in\left\{\mathcal{T}^\alpha\right\}}P^\alpha\!
    \left(\mathcal{T}^\alpha\right)f^\alpha\!\left(\mathcal{T}^\alpha;D\!\left(S\right)\right).
\end{equation}

In the following, we will show that a MLD as defined
in using \cref{eq:MLDgeneral} maximizes the sample average $\langle\langle
f^\alpha_D\rangle\rangle$:
\begin{equation}
	\begin{split}
		&\sup_D\left[\langle\langle f^\alpha_D\rangle\rangle_\alpha\right]
        \\
		=&\sup_D\left[\sum_{\mathcal{T}^\alpha\in\left\{\mathcal{T}^\alpha\right\}}P^\alpha\!
        \left(\mathcal{T}^\alpha\right)f^\alpha\!\left(\mathcal{T}^\alpha;D\!\left(S\right)\right)\right].
	\end{split}
\end{equation}
It is important to note that here we consider all possible decoders of
the form $D\!\left(S\right)$ which take stabilizer measurements $S$
as an input and return a correction string. Now we realize that the
trajectories $\mathcal{T}^\alpha$ can be grouped into equivalence classes
$\left[\mathcal{T}\right]=S\!\left(\mathcal{T}^\alpha\right)$ via their
respective stabilizers $S$. This allows us to rewrite the sum by first
considering all possible stabilizers $S$ and for each one of them summing
over all trajectories:
\begin{align}
	&=\sup_D\!\left[\sum_S\sum_{\mathcal{T}^\alpha\in S}P^\alpha\!\left(\mathcal{T}^\alpha\right)f^\alpha\!\left(\mathcal{T}^\alpha;D\!\left(S\right)\right)\right].
\end{align}

In the following, we use \cref{eq:MLDProofMaximum} to take the supremum into
the outer sum. We can do this because we just require the decoder $D$ to
be any function that takes stabilizers $S$ as input and returns correction
strings. Thus, we find
\begin{equation*}
	=\sum_S\sup_D\!\left[\sum_{\mathcal{T}^\alpha\in S}P^\alpha\!
    \left(\mathcal{T}^\alpha\right)f^\alpha\!
        \left(\mathcal{T}^\alpha;D\!\left(S\right)\right)\right].
\end{equation*}
As the decoder $D\!\left(S\right)$ is just a function that returns a
correction string $C$ depending on the input $S$ and we just fixed $S$,
we can now just write the supremum as
\begin{equation*}
	=\sum_S\sup_C\!\left[\sum_{\mathcal{T}^\alpha\in S}P^\alpha\!
    \left(\mathcal{T}^\alpha\right)f^\alpha\!\left(\mathcal{T}^\alpha;C\right)\right].
\end{equation*}
We can now consider our definition of a MLD in \cref{eq:MLDgeneral} and
rewrite as follows: \emph{For a given $S$, choose the correction string $C_i$
such that}
\begin{equation}
	P^\alpha_{f^\alpha}\!\left(C_i\vert S\right)\cdot P\!\left(S^p\right)
    =\sum_{\mathcal{T}^\alpha\in S}P^\alpha\!
    \left(\mathcal{T}^\alpha\right)f^\alpha\!\left(\mathcal{T}^\alpha;C_i\right)
\end{equation}
\emph{is maximized.} Here the sum does only consider
trajectories $\mathcal{T}^\alpha\in S$. We can thus use the MLD
$D_\sub{ML}^\alpha\!\left(S\right)$ to finalize our calculation:
\begin{equation}
	\begin{split}
		&\sup_D\left[\langle\langle f^\alpha_D\rangle\rangle_\alpha\right]
        \\
		=&\sum_S\sum_{\mathcal{T}^\alpha\in S}P^\alpha\!\left(\mathcal{T}^\alpha\right)f^\alpha
        \left(\mathcal{T^\alpha;D_\sub{ML}^\alpha\!\left(S\right)}\right)\\
		=&\langle\langle f^\alpha_{D_\sub{ML}^\alpha}\rangle\rangle_\alpha.
	\end{split}
\end{equation}
From \cref{app:simulateClassical}, we know that $\langle\langle
f^\sub{bi}_D\rangle\rangle_\text{bi}=\langle\langle
f^\sub{qm}_D\rangle\rangle_\text{qm}$. Thus, we find the following connection
between the classical and the quantum MLD. It is
\begin{equation}
	\begin{split}
		&\sup_D\left[\langle\langle f^\sub{qm}_D\rangle\rangle_\text{qm}\right]
        \\
		=&\langle\langle f^\sub{qm}_{D^\sub{qm}_\sub{ML}}\rangle\rangle_\text{qm}
	\end{split}
\end{equation}
and
\begin{equation}
	\begin{split}
		&\sup_D\left[\langle\langle f^\sub{bi}_D\rangle\rangle_\text{bi}\right]
        \\
		=&\langle\langle f^\sub{bi}_{D^\sub{bi}_\sub{ML}}\rangle\rangle_\text{bi}.
	\end{split}
\end{equation}
Using \cref{app:simulateClassical}, we find
\begin{align}
	\langle\langle f^\sub{qm}_{D^\sub{qm}_\sub{ML}}\rangle\rangle_\text{qm}
    &=\sup_D\!\left[\langle\langle f^\sub{qm}_D\rangle\rangle_\text{qm}\right]
    \\
	&=\sup_D\!\left[\langle\langle f^\sub{bi}_D\rangle\rangle_\text{bi}\right]
    \\
	&=\langle\langle f^\sub{bi}_{D^\sub{bi}_\sub{ML}}\rangle\rangle_\text{bi}
    \\
	&=\langle\langle f^\sub{qm}_{D^\sub{bi}_\sub{ML}}\rangle\rangle_\text{qm}.
\end{align}
This proves that the classical MLD yields the exact same results as the
quantum mechanical MLD.\hfill$\blacksquare$

\section{Proof of the Sampling in \cref{eq:P(Tbi;Ep)}}
\label{sec:pseudoclassicalSampling}

Here we show a short proof for the sampling discussed in
\cref{eq:P(Tbi;Ep)}. We do so by inserting \cref{eq:P(Tbi;Ep)} in
\cref{eq:pseudoclassicalRelation} to obtain \cref{eq:P(Tbi)}.

Consider the term
\begin{equation}
	\frac{1}{P^\sub{bi}\!\left(S^p\right)}
    \sum_{E^p_{\vert\mathcal{T}^\sub{bi}}}P^\sub{bi}\!\left(\mathcal{T}^\sub{bi};E^p\right),
\end{equation}
where we take the sum over all error patterns $E^p$ that contain the bit
flip pattern $\flpat$ which is part of $\mathcal{T}^\sub{bi}$. Inserting
\cref{eq:P(Tbi;Ep)} yields
\begin{align}
	&=\sum_{E^p_{\vert\mathcal{T}^\sub{bi}}}p^{\left|E^p\right|}\cdot
    \left(1-p\right)^{LT-\left|E^p\right|}\cdot\left(\frac{1}{2}\right)^{\left|E^p\right|}
    \\
	&=\sum_{E^p_{\vert\mathcal{T}^\sub{bi}}}
    \left(\frac{p}{2}\right)^{\left|E^p\right|}\cdot\left(1-p\right)^{LT-\left|E^p\right|}
    \\
	&=\left(\frac{p}{2}\right)^{\left|\flpat\right|}
    \left(1-p\right)^{LT-\left|\flpat\right|}\sum_{E^p_{\vert\mathcal{T}^\sub{bi}}}
    \left(\frac{p}{2\left(1-p\right)}\right)^{\left|E^p\right|-\left|\flpat\right|}.
\end{align}
Now consider the summation over all possible patterns $E^p$
consistent with the trajectory $\mathcal{T}^\sub{bi}$. This
is equivalent to a summation over all virtual patterns
$E^p_\text{v}:=E^p\setminus\flpat$. This pattern is constructed such that
$0\le\left|E^p_\text{v}\right|=\left|E^p\right|-\left|\flpat\right|\le
LT-\left|\flpat\right|$ and allows us to rewrite the sampling as
\begin{equation}
	=\left(\frac{p}{2}\right)^{\left|\flpat\right|}
    \left(1-p\right)^{LT-\left|\flpat\right|}\sum_{E^p_\text{v}}
    \left(\frac{p}{2\left(1-p\right)}\right)^{\left|E^p_\text{v}\right|}.
\end{equation}
It becomes clear now that sampling over all possible virtual patterns
$E^p_\text{v}$ is equivalent to choosing all possible subsets of a set with
$LT-\left|\flpat\right|$ elements. This can be formulated as
\begin{align}
	\begin{split}
		=&\left(\frac{p}{2}\right)^{\left|\flpat\right|}\left(1-p\right)^{LT-\left|\flpat\right|}
        \\
		&\cdot\sum_{k=0}^{LT-\left|\flpat\right|}
        \begin{pmatrix}LT-\left|\flpat\right|\\k\end{pmatrix}
        \left(\frac{p}{2\left(1-p\right)}\right)^k.
	\end{split}
    \\
	\begin{split}
		=&\left(\frac{p}{2}\right)^{LT}
        \\
		&\cdot\sum_{k=0}^{LT-\left|\flpat\right|}
        \begin{pmatrix}LT-\left|\flpat\right|\\k\end{pmatrix}
        1^k\left(\frac{2\left(1-p\right)}{p}\right)^{\left(LT-\left|\flpat\right|\right)-k}.
	\end{split}
\end{align}
Using the binomial theorem, we find
\begin{align}
	&=\left(\frac{p}{2}\right)^{LT}\cdot\left(1+\frac{2\left(1-p\right)}{p}\right)^{LT-\left|\flpat\right|}
    \\
	&=\left(\frac{p}{2}\right)^{\left|\flpat\right|}\cdot\left(1-\frac{p}{2}\right)^{LT-\left|\flpat\right|}
    \\
	&=\frac{P^\sub{bi}\!\left(\mathcal{T}^\sub{bi}\right)}{P^\sub{bi}\!\left(S^p\right)}.
\end{align}
This proves the relation in \cref{eq:pseudoclassicalRelation} and thus proves
that the sampling in \cref{eq:P(Tbi;Ep)} is correct.\hfill$\blacksquare$

\section{Thermodynamic limit of the MWPM decoder for $q=0$}
\label{app:finitesize}

\begin{figure}[tb]
	\centering \includegraphics[width=\linewidth]{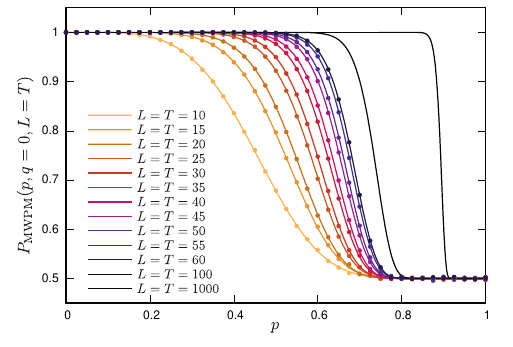}
    \caption{%
        \CaptionMark{Analytical investigation of the MWPM decoder for the
        case $q=0$.}
        The lines in this plot show the analytical results of the MWPM decoder
        in the case $q=0$. The data points are gained from simulations with
        $100\,000$ samples.
    } 
    \label{fig:MWPMq0}
\end{figure}

Here we want to shortly discuss the performance of the MWPM decoder for $q=0$
in the thermodynamic limit. For this special case, analytical calculations
can be made.

For $q=0$, all stabilizers are measured and the lattice produced by the
MWPM decoder as shown in \cref{fig:MWPMa}~(b) only features unconnected
horizontal lines. Essentially, the MWPM decoder is now a MVD as discussed
in \cref{sec:MVD}. Now if in a classical system of bits the number of flips
in a single step randomly surpasses $\nicefrac{L}{2}$ in a time step (which
has a nonzero likelihood in finite systems), the decoder will match the nodes
falsely and fail in this time step. If this happens in an even number of time
steps, the decoder yields the correct result after the last step, otherwise
it fails. Therefore, we can calculate the decoding probability analytically via
\begin{subequations}
	\begin{align}
		\begin{split}
			&P_{D_\sub{MWPM}}\!\left(p,q=0;L,T\right)\\
			&=\sum_{t=0,t\;\text{even}}^{T}\left[\begin{pmatrix}T\\ t\end{pmatrix}\cdot\left(P\!\left(L,p\right)\right)^t\cdot\left(1-P\!\left(L,p\right)\right)^{T-t}\right],
		\end{split}
		\\
		\begin{split}
			&P\!\left(L,p\right)\\
			&=\sum_{b=\lceil\nicefrac{L}{2}\rceil}^{L}\begin{pmatrix}L\\ b\end{pmatrix}\left(\frac{p}{2}\right)^b\cdot\left(1-\frac{p}{2}\right)^{L-b}\cdot
			\begin{cases}
				\frac{1}{2} & \text{if }b=\frac{L}{2}\\
				1 & \text{else}
			\end{cases}.
		\end{split}
	\end{align}
\end{subequations}
%
The function $P\!\left(L,p\right)$ calculates the probability of more
than $\nicefrac{L}{2}$ bits being flipped in a single time step. If exactly
$\nicefrac{L}{2}$ bits are flipped, the MWPM decoder randomly guesses correctly
or falsely. Therefore, we multiply this probability by $\nicefrac{1}{2}$.

The analytical function was plotted for different system sizes in
\cref{fig:MWPMq0}. The dots in the plot represent simulation data. They
clearly fit the theory. As can be seen, the decoding probability exhibits
a strong finite-size effect but in the thermodynamic limit $L=T\to\infty$,
the decoding probability slowly approaches $1$ for all $p<1$.

\end{document}